\documentclass{article}

\usepackage{amsmath}
\usepackage{amssymb}
\usepackage{amsthm}
\usepackage{graphicx}
\usepackage{verbatim}
\usepackage{epsfig}
\usepackage{color}
\usepackage{tikz}
\usepackage{url}
\usepackage[caption=false,font=footnotesize]{subfig}
\usepackage[english]{babel}
\usepackage{algorithm}
\usepackage[noend]{algorithmic}
\usepackage{booktabs,multirow}
\usepackage{cleveref}
\usepackage{xspace}
\usepackage[utf8]{inputenc}
\usepackage[T1]{fontenc}
\usepackage{authblk}

\usetikzlibrary{decorations.pathreplacing, calc, positioning}

\newtheorem{theorem}{Theorem}[section]
\newtheorem{lemma}[theorem]{Lemma}
\newtheorem{proposition}[theorem]{Proposition}

\DeclareMathOperator*{\argmax}{argmax}

\begin{document}

\title{Scheduling on Two Types of Resources: a Survey}

\date{}

\author[1]{Olivier Beaumont}
\author[2]{Louis-Claude Canon}
\author[1]{Lionel Eyraud-Dubois}
\author[3]{Giorgio Lucarelli}
\author[4]{Loris Marchal}
\author[5]{Cl\'ement Mommessin}
\author[6]{Bertrand Simon}
\author[5]{Denis Trystram}

\affil[1]{INRIA Bordeaux, France}
\affil[2]{FEMTO-ST, Université de Bourgigne Franche-Comté, France}
\affil[3]{LCOMS, University of Lorraine, Metz, France}
\affil[4]{CNRS, Univ. Lyon, LIP, France}
\affil[5]{Univ. Grenoble Alpes, CNRS, INRIA, Grenoble INP, LIG, France}
\affil[6]{University of Bremen}

\maketitle

\begin{abstract}
The evolution in the design of modern parallel platforms leads to revisit
the scheduling jobs on distributed heterogeneous resources.
The goal of this survey is to present the main existing algorithms
to classify them based on their underlying principles and to propose 
unified
implementations to enable their fair comparison, both in terms of 
running time and
quality of schedules, on a large set of common benchmarks that we made available for the community.
Beyond this comparison, our goal is also to understand the main 
difficulties that
heterogeneity conveys and the shared principles that guide the design
of efficient algorithms.
\end{abstract}

\section{Introduction}\label{sec:intro}

A key ingredient of any computing system is the scheduler, which is responsible for handling the tasks submitted by the users and the computing resources.
Specifically, the scheduling algorithm has to decide which task to execute first, when to start its execution, and where to allocate it (i.e., which resources to use).
Due to the importance of these decisions, the efficiency of the scheduler is crucial for the performance of the whole system.

Scheduling is a well understood problem in the
context of homogeneous platforms composed of identical resources,
since there exist both efficient theoretical approximation
algorithms~\cite{Drozdowski:2009:SPP} and their practical counterpart
implementations in actual batch schedulers, such as SLURM~\cite{SLURM}
or Torque~\cite{Torque}. However, the case of heterogeneous resources,
which is crucial in practice due to the evolution of architectures, is
not so well understood and it has been the focus of a vast literature
in recent years~\cite{HEFT, MinMin, Boeres, lookaheadHEFT}.
The purpose of this survey is to make an attempt to
unify the results and to provide a clear review of existing solutions.
In this context, our goals are: (i) to understand the intrinsic difficulties introduced by heterogeneity, (ii) to classify the different approaches to deal with heterogeneity, and (iii) to provide a comprehensive way to evaluate the performance of reviewed algorithms, both in theory and experimentally and both in terms of quality of produced results and in terms of running time.

A good example to understand the difference between homogeneous
and heterogeneous cases is the well-known greedy List Scheduling
algorithm~\cite{Graham69}, which minimizes the maximum completion time
of a parallel application (i.e., \emph{makespan}) and which is arbitrarily
bad in heterogeneous platforms. Adapting List Scheduling algorithms in
the context of heterogeneous resources to obtain low cost algorithms
whose performance can be theoretically assessed is not easy. As we
will show in the paper, it is possible to design variants of List Scheduling,
but at the price of putting the
emphasis on the allocation on the right set of resources. The
analysis of the complexity induced by heterogeneity leads to the
identification of the main scientific locks, and then to propose a
two-phase approach for designing efficient scheduling algorithms. In
this survey, we revisit existing algorithms within the same unified
framework and we report a common experimental campaign for comparing
them.  The associated benchmark and simulation framework will be of great interest for
further related studies.\\

We consider both off-line and on-line settings and target the most
popular objective, which is the minimization of the makespan. In the
off-line setting, the whole set of tasks is known in advance, while in the
on-line setting, tasks arrive one by one and there is no a priori knowledge
of the upcoming tasks. The on-line setting is more difficult but it is
of particular interest in the context of the large scale heterogeneous
platforms that we target. Indeed, in the context of heterogeneous platforms, programmers of parallel applications, including for regular applications such as dense linear algebra, rely of runtime dynamic systems. These schedulers, such as Quark~\cite{YarKhan:2011:Quark:Manual},
ParSeC~\cite{parsec}, StarSs~\cite{starss} and StarPU~\cite{AugThiNamWac11CCPE}, make all their allocation and scheduling decisions at runtime. These on-line decisions are based on the state of the platform, the set of available (ready) tasks, and possibly on static pre-allocation strategies and tasks priorities that have been computed offline.

In order to model the performance of heterogeneous resources, we
consider a fully unrelated model where processing times are provided
for each (type of) task and for each (type of) resource.  We focus on
the case of two types of resources since it corresponds to the widely
spread setting of machines consisting of CPU multicores and GPUs
(extensions are discussed in \Cref{sec:extensions}).  We study
parallel applications composed of tasks with or without precedence
relations~\cite{BookCosnardTrystram} and whose execution times are known.
\bigskip

The roadmap of the paper is as follows: the general model is described in
\Cref{sec:definitions}.  The methodology used throughout this paper
to obtain a unified way to present, establish and theoretically compare
 the results is presented in \Cref{sec:methodology}.
The theoretical results, in particular the lower bounds for both the
off-line and the on-line cases, are given in \Cref{sec:theory}.
\Cref{sec:independent} is dedicated to the special case of independent
tasks, which is inherently easier than the case with dependencies.
This case, due to its interest in the context of runtime
schedulers, lead to the design of low cost algorithms that nevertheless
achieve guaranteed approximation ratios.  The case with dependencies
is considered in \Cref{sec:prec}, where we discuss both the classical
heuristics of the literature such as HEFT~\cite{HEFT2002} and review
approximation algorithms with constant approximation ratios in the
off-line case (\Cref{ssec:precOFF}). \Cref{ssec:precON} also provides
approximation algorithms in the on-line case. \Cref{sec:independent}
and \Cref{sec:prec} therefore establish, in a unified theoretical
framework, a comprehensive survey of the recent results of the
literature and enable to understand the intrinsic difficulties
introduced respectively by resource heterogeneity and by the on-line
setting. On the other hand, the goal of \Cref{sec:experiments} is to
provide a fair experimental comparison framework of all reviewed algorithms,
using a large set of benchmark problems and platform
architectures. The experiments aim at comparing the algorithms both in
terms of their actual scheduling performance and in terms of their
running times. Then, extensions to more types of resources are
discused in \Cref{sec:extensions}. Finally, we conclude with a
synthesis and discussion on both theoretical and practical results in
\Cref{sec:conclusion}.

\section{Definitions and Notations}\label{sec:definitions}

As mentioned in the introduction, we focus in this paper on the case of two types of resources, which is of special practical interest.
We consider a set $\mathcal{T}$ of $n$ sequential tasks that are to be scheduled on a platform composed of $m$ identical processors of type 1 and $k$ identical processors of type 2. Without loss of generality, we denote by CPU (resp. GPU) a processor of type 1 (resp. type 2), and we assume that $m \geq k$.
The processing of a task requires a different amount of time when performed on a CPU or on a GPU.
Let $\overline{p_j}$ (resp. $\underline{p_j}$) denote the processing time of task $j$ when processed on a CPU (resp. GPU) and let $\alpha_j = \frac{\overline{p_j}}{\underline{p_j}}$ be the acceleration factor of $j$.
Note that despite its name, this acceleration factor may well be smaller than~1 in the case of a task that runs faster on a CPU.

Moreover, given a schedule $S$, let $C_j$ denote the completion time of task $j$ and $x_j$ be the binary variable that indicates where $j$ is processed ($x_j=1$ when $j$ is processed on a CPU, and $x_j=0$ otherwise).
Therefore, $x_j$ denotes the allocation of task $j$ to a processor type.

At last, let us denote by $\overline{W}$ the overall workload on CPUs for the schedule $S$, given by $\overline{W} = \sum_{j \in \mathcal{T}}{x_j \overline{p_j}}$. Similarly, the overall workload on all GPUs is given by $\underline{W} = \sum_{j \in \mathcal{T}}{(1-x_j) \underline{p_j}}$.

If the application tasks are linked by priority relationships, the set of tasks $\mathcal{T}$ is seen as a directed acyclic graph $G=(V,E)$, whose vertices correspond to the tasks and arcs correspond to the precedence relationships between the tasks.
In any feasible schedule, for each arc $(i,j) \in E$, $j$ cannot start its execution before the completion of $i$.
In this case, $i$ is said to be a \emph{predecessor} of $j$ and $\Gamma^-(j)$ denotes the set of all predecessors of $j$.
Similarly, $j$ is said to be a \emph{successor} of $i$ and $\Gamma^+(i)$ denotes the set of all successors of $i$.
A \emph{descendant} of $j$ is a task $i$ for which there exists a path from $j$ to $i$ in $G$.

Given a schedule $S$, let us denote by $CP$ the critical path of the schedule, i.e., the longest weighted path, where the weights correspond to processing times, between any two tasks of $G$.
Let us also define the \emph{bottom-level} of a task $j$ as the longest weighted path between $j$ and any of its descendants, the processing time of $j$ being excluded, as introduced by Yang and Gerasoulis~\cite{GerasoulisDSC}.
Note that these two definitions are associated to a specific schedule $S$, where the allocation of tasks on either a CPU or a GPU is done, so that the processing times of all tasks are known.
On the other hand, if the allocation of the tasks onto resources is not yet determined, each task comes with two possible processing times and the closest notion to the critical path is the \emph{upward rank} of tasks used in the HEFT algorithm~\cite{HEFT2002} (and presented in~\Cref{sssec:HEFT}).

\medskip 
In this context, the goal is to build a feasible and non-preemptive schedule of minimum makespan, denoted by $C_{max}$, that satisfies all precedence relations between tasks, if any. In other words, we are looking for a schedule where the execution of any task cannot be interrupted and that minimizes the completion time of the last finishing task, i.e., $C_{max} = \max_{j \in \mathcal{T}}C_j$.

\medskip 
Using the three-field notation for scheduling problems introduced by Graham~\cite{3fieldsGraham}, these two problems, with independent tasks and tasks linked by precedence constraints, can be denoted respectively as $(Pm,Pk) \mid \mid C_{max}$ and $(Pm,Pk) \mid prec \mid C_{max}$.
These problems are harder to solve than $P \mid \mid C_{max}$ and $P \mid prec \mid C_{max}$, which are known to be NP-hard~\cite{garey} but admits Polynomial Time Approximation Schemes (PTAS)~\cite{approxSchemes, dualApprox}.
However, they are easier to solve than scheduling problems on
unrelated machines ($R \mid \mid C_{max}$ and $R\mid prec \mid
C_{max}$) since we consider only two types of resources.
Although several approximation algorithms and PTAS have been proposed
for these scheduling problems on unrelated
machines~\cite{lenstra1990a,shmoys1993a,shchepin2005a,gairing2007a},
their costs make them impractical for runtime schedulers.
Moreover, PTAS have been proposed for the problems we tackle when
considering a constant number $K$ of processor
types~\cite{bonifaci2012a,gehrke2016a}, but the cost of these
approaches is however prohibitive even with $K=2$.
Finally, the problems we consider reduce to the scheduling problems on uniformly related machines ($Q \mid \mid C_{max}$ and $Q\mid prec \mid C_{max}$) if all tasks have the same acceleration factor, i.e., if $\alpha_j = \alpha,~ \forall j \in \mathcal{T}$, and for which there exists a
$\log(p)$-approximation in presence of precedences~\cite{chudak1999}.

\medskip
The above scheduling problems are said to be in the \textit{off-line} context when the set of tasks to be scheduled and their processing times are known in advance.
In this work, we also consider the \textit{clairvoyant on-line} context as defined by Leung~\cite[Chapter 15]{leung2004handbook}.
In this case, we assume that: (i) a task arrives in the system when it becomes ready, i.e., when all its predecessors have been processed, (ii) when a task arrives, its processing time on any type of resource is known to the scheduler, and (iii) the scheduler must schedule a task immediately and irrevocably upon its arrival, without knowing anything on the upcoming tasks.
If multiple tasks become ready at the same time, we consider that they arrive in the system in any order.
This is the case for independent tasks that all become ready at time~0.

\section{Preliminaries and Methodology}\label{sec:methodology}

\subsection{Preliminaries}\label{ssec:preliminaries}

Since our aim is to design scheduling algorithms with performance guarantees, we rely on the notions of \emph{approximation ratio} and \emph{competitive ratio} presented by Hochbaum~\cite{Hochbaum}.
The approximation ratio (resp. competitive ratio) $\rho_A$ of an off-line (resp. on-line) algorithm $A$ is defined as the maximum, defined over all possible instances $I$ of the considered problem, of the ratio $\frac{C_{max}(I)}{C_{max}^*(I)}$, where $C_{max}(I)$ denotes the makespan of $A$ on the instance $I$ and $C_{max}^*(I)$ is the optimal \emph{off-line} makespan on the instance $I$.
In this section, we concentrate on the homogeneous setting, where there are $m$ resources of the same type only.
Therefore, for the sake of simplicity we will denote by $W$ the overall work and by $p_{max}$ the maximal completion time of a task.

\subsubsection{List Scheduling approximation ratio without precedences}
\label{sec:list-sched-appr}

One of the first results in Scheduling Theory concerns the \emph{List Scheduling} algorithm introduced by Graham~\cite{Graham69} for the problem of scheduling a list of independent tasks on $m$ identical parallel machines ($P \mid \mid C_{max}$).
List Scheduling is built as follows: whenever a processor becomes idle, it processes the first task in the list of still unprocessed tasks.
Therefore, the time complexity of this procedure is $O(n\log(m))$.

It is easy to establish that this algorithm achieves an approximation ratio of at most 2: consider a schedule produced by List Scheduling of makespan $C_{max}$.
We can partition the time interval $[0, C_{max})$ of this schedule into two intervals $[0, t)$ and $[t, C_{max})$ such that $t$ is the earliest time when at least one processor becomes idle, as depicted on the left part of \Cref{fig:proofLS}.
The length of the first time interval (before $t$) can be upper bounded by the average load of a processor ($\frac{{W}}{m}$), which is itself bounded by the optimal makespan $C_{max}^*$.
To bound the length of the second interval, we can notice that no task can start its execution strictly after time $t$, otherwise it would have started on a processor that is idle at time $t$.
Thus, we can upper bound the length of the second interval by the longest execution time of any task (${p_{max}}$), which is also bounded by $C_{max}^*$.

We therefore obtain the following bound, $$ C_{max} \leq \frac{{W}}{m} + {p_{max}} \leq 2 \cdot C_{max}^*,$$ which concludes the proof.
By carefully evaluating the contribution of the work of the last ending task in the first phase, this upper bound can be further lowered to $(2 - \frac{1}{m}) C_{max}^*$.

Let us notice that, since all tasks are independent, it is possible to re-order the list of tasks following a given policy.
For example, re-ordering the tasks in the list by decreasing processing time leads to long tasks being executed first in the schedule, thus leading to a decrease of the length of the last time interval $[t, C_{max})$.
This policy is denoted as \emph{Largest Processing Time} (LPT) and improves the approximation ratio of List Scheduling to $\frac{4}{3} - \frac{1}{3m}$~\cite{Graham69}.

\subsubsection{List Scheduling approximation ratio with precedences}

If precedence relations between tasks are considered, no global re-ordering of the list of tasks is possible and List Scheduling works as follows.
Whenever a processor becomes idle, it scans the list of tasks and processes the first ready task that it finds.
If no task is ready to be processed (due to precedence constraints), the processor waits until a running task is completed by another processor and then it re-tries to schedule a newly ready task.
Therefore, a resource is idle at time $t$ if and only if there is no ready task in the list at time $t$.
For each task, all successors are analyzed to determine which tasks
become ready. Moreover, determining the next idle processor once a
task has been scheduled requires $\log(m)$ operations.
Therefore, the time complexity of this approach is $O(n\log(m)+|E|)$
where $|E|$ is the number of dependencies.

\begin{theorem}\label{thm:GrahamLS}
List Scheduling for tasks with precedence constraints is a $(2-\frac{1}{m})$-approximation algorithm.
\end{theorem}

\begin{figure}
\begin{center}
\includegraphics[scale=.12]{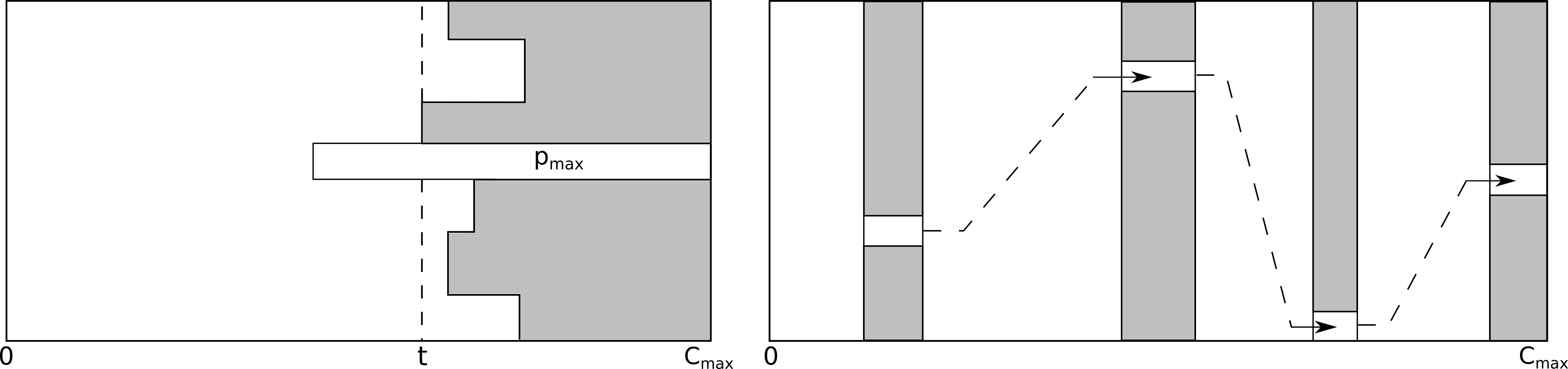}
\caption{Left: Partition of an output schedule of List Scheduling for independent tasks.
Right: Construction of a chain of tasks based on the idle time intervals, denoted by grey areas.}\label{fig:proofLS}
\end{center}
\end{figure}

The proof~\cite{Graham69} follows the same principle as for the case of independent tasks by replacing the bound on ${p_{max}}$ by a bound on the sum of idle times in the schedule and by using a geometrical argument.
Indeed, let us notice that $C_{max} = \frac{{W}}{m} + \frac{S_{idle}}{m}$, where $S_{idle}$ is the sum of idle times on all processors.
The first term $\frac{{W}}{m}$ is the same as before and can be upper bounded by $C_{max}^*$.
For the second term $\frac{S_{idle}}{m}$, let us consider the last finishing task $j$ in the schedule.
This task must be a successor of a task executed during the previous interval where there is an idle time, otherwise the algorithm would have scheduled $j$ during this idle time. Thus, we can iteratively build a chain of tasks (of total duration $L$) that are being processed during each idle time interval of the schedule, as depicted on the right part of \Cref{fig:proofLS}.
During the time intervals where there are idle times, there are at most $m-1$ idle processors, leading to $S_{idle} \leq (m-1) L$.
Since the length $L$ of this chain of tasks is smaller than the length of the critical path, we obtain $S_{idle} \leq (m-1) |CP| \leq (m-1) C_{max}^*$.

Thus, using the upper bounds on both terms, we obtain the following bound
\[ C_{max} = \frac{{W}}{m} + \frac{S_{idle}}{m} \leq C_{max}^* + \frac{m-1}{m} C_{max}^* = (2-\frac{1}{m}) C_{max}^*, \]
which concludes the proof.

\medskip
In 2011, Svensson~\cite{Svensson} showed that this result was the best possible bound by proving that, if we consider a variant of the unique games conjecture~\cite{UGCvariant}, it is NP-hard to approximate this scheduling problem ($P \mid prec \mid C_{max}$) within a factor smaller than 2, even in presence of unit processing times.

\subsubsection{Dual approximation}
\label{sec:dual-approximation}

Dual approximation~\cite{dualApprox} is a useful technique for designing effective approximation algorithms, which is used in some of the scheduling strategies discussed in this paper.
In order to design a $\rho$-dual approximation algorithm for a general scheduling problem, the process starts from an initial guess $\lambda$. Then, it either determines that makespan $\lambda$ is not achievable by any algorithm, or it outputs an actual schedule whose length is at most $\rho \lambda$. Since above process can be applied with any value $\lambda$, it can be incorporated into a binary search process, starting from a lower bound $B_{min}$ and an upper bound $B_{max}$ on the optimal makespan. At each step of the binary search, if there is no schedule with makespan $\lambda$, then $\lambda$ becomes the new lower bound and if there exists a schedule with makespan $\rho \lambda$, then $\rho \lambda$ becomes the new upper bound.

We iterate binary search until the gap between both bounds becomes smaller than a desired precision threshold $\epsilon$ and the number of steps is upper bounded by $\log_{2}(\frac{B_{max} - B_{min}}{\epsilon})$.
Therefore, it is possible to turn a $\rho$-dual approximation algorithm into a $\rho(1 + \epsilon)$-approximation algorithm with the same computational complexity bound.

\subsection{Methodology}\label{ssec:methodo}

We are interested in studying generic strategies generating schedules for hybrid computing systems consisting of CPUs and GPUs that behave well \textit{in practice}, i.e., when they are actually used to run parallel applications.
As stated in~\Cref{sec:intro}, we will consider several cases, namely both independent tasks and tasks linked by precedence relations, and in both the off-line and on-line settings.

Our purpose is not only to survey or classify existing results, but to highlight the main ideas behind these algorithms, to emphasize their differences and to analyze their actual performance.
More precisely, our methodology for each algorithm is to start by giving an insight of the key ideas and then to present the main stages of the algorithms, where the notations and the writing styles have been normalized to allow the reader to perform an easy comparison.

In order to prove performance guarantees, we provide in Section~\ref{sec:theory} a set of lower bounds corresponding to the different settings and that will be used throughout the paper.
As we will see, some of these bounds can be used in the different cases.  Then, for each presented algorithm in a specific setting, we establish its worst case performance against the associated lower bounds (\Cref{sec:independent} and \Cref{sec:prec}).

In order to provide another fair comparison between the different algorithms, we rely on experimental results using realistic benchmarks, capturing both the characteristics of hybrid platforms and the characteristics of the applications they process.
For this purpose, we develop a shared experimental framework under which all algorithms are evaluated, coded in a standardized way (under the same language), on the same simulation testbed and a datasets built from actual execution traces.

\section{Lower Bounds}\label{sec:theory}

This section gathers the lower bounds from the literature for the hybrid scheduling problem, first in the off-line context and then in the on-line context.
In each case, we distinguish between applications consisting of independent tasks and those with precedence constraints.

\subsection{Lower Bounds on the Optimal Makespan}\label{ssec:makespanLB}
\subsubsection{With independent tasks}\label{sssec:indepOffLB}

Let us consider the problem defined in Section~\ref{sec:definitions}.
It is possible to derive a set of lower bounds on the best achievable makespan by any valid schedule.
A first trivial lower bound states that no schedule can be shorter than the longest task, i.e.,
\[ \max_{j \in \mathcal{T}}\min(\overline{p_j}, \underline{p_j}) \leq C_{max}^* \]

The second lower bound to evaluate the performance of scheduling algorithms for independent tasks is called \emph{Area Bound} and it is based on the maximal amount of work that the different resources can perform in a given amount of time.
Let us first remark that any schedule $S$ uses at most $m$ CPUs and $k$ GPUs for a duration at most $C_{max}$, so that the following constraints hold true:
\[ \sum_{j \in \mathcal{T}} x_j \overline{p_j} = \overline{W} \leq m\cdot C_{max} \quad \text{and} \quad
\sum_{j \in \mathcal{T}} (1-x_j) \underline{p_j} = \underline{W} \leq
k\cdot C_{max} \text{.}\]

Thus, the following Linear Program (LP) $LP_{\text{Area}}$ can be seen as a relaxation of the scheduling problem, where the set of $m$ CPUs (resp. the set of $k$ GPUs) is considered as a whole.

\begin{align*}
   \text{minimize } \quad & ~C_{LP} \\ \text{subject to:} \quad &
   \sum_{j \in \mathcal{T}} \overline{p_j} x_j \leq m\cdot C_{LP} \\ &
   \sum_{j \in \mathcal{T}} \underline{p_j} (1-x_j) \leq k\cdot C_{LP}
   \\ & x_j \in [0, 1], \forall j \in \mathcal{T}
\end{align*}

Since any schedule $S$ of makespan $C_{max}$ can be transformed into a solution of $LP_{\text{Area}}$ with objective $C_{LP}$, then the optimal solution of $LP_{\text{Area}}$ holds as a lower bound of the optimal makespan $C_{\text{max}}^*$.
It can be proved~\cite{HeteroPrio} that this optimal solution has a specific structure, where all the tasks with $x_j = 1$ have an acceleration factor lower than all the tasks with $x_j = 0$.
In such case the optimal solution of the LP can in fact be obtained using a greedy algorithm which sorts tasks by non-decreasing acceleration factors and then assigns at the same rate tasks at the beginning of the list to CPUs, and tasks at the end of the list to GPUs, until only one task remains, which is adequately split so that the total time on CPUs and GPUs is the same.
This lower bound and its specific structure is also described by Canon et al.~\cite[Theorem 5]{BalancedEuropar}.
Its closed form expression considers the pivot task $i$ (the one split between CPUs and GPUs) and results in the exact same lower bound.

\begin{theorem}\label{th:lower-bound}
  Assume tasks are sorted by non-decreasing acceleration factor ($\alpha_i \leq \alpha_j$ for $i<j$). Let $i$ denote the task such that
  $$ \frac{1}{m}\sum_{j\leq i} \overline{p_j} \geq
  \frac{1}{k}\sum_{j>i} \underline{p_j} \quad \textnormal{and} \quad
  \frac{1}{m}\sum_{j< i} \overline{p_j} \leq \frac{1}{k}\sum_{j\geq i}
  \underline{p_j}.
  $$ 
Then, 
  $$ \textnormal{LB} =\frac{\displaystyle \underline{p_i} \sum_{j<i}
    \overline{p_j} + \overline{p_i}\sum_{j>i} \underline{p_j} +
    \overline{p_i} \underline{p_i}}{k \overline{p_i} + m
    \underline{p_i}}.
  $$
is a lower bound on the optimal makespan.
\end{theorem}

In the (heterogeneous) case of two types of resources, many approximation results that hold true in the homogeneous model cannot be extended.
For example, the well-known List Scheduling algorithm, which has been proved to be a 2-approximation by Graham~\cite{Graham69} as shown in Section~\ref{sec:methodology}, fails to achieve a constant approximation ratio in the heterogeneous setting.
Indeed, for any integer $l > 1$, let us consider an instance with exactly one resource of each type (i.e., $m=k=1$), and two tasks with $\overline{p_j}=M \gg 1$ and $\underline{p_j} = 1$.
Clearly, an optimal schedule in this case allocates both tasks on the GPU, and thus achieves a makespan of $2$.
However, any List Scheduling algorithm does not let the CPU idle at time $0$, and therefore achieves a makespan $M$.
Since $M$ can be arbitrarily large, this proves that a List Scheduling algorithm cannot achieve a constant approximation ratio.


\subsubsection{With precedence constraints} \label{sssec:precLB}

Considering tasks with precedence constraints, an extension of $LP_{area}$ (\Cref{sssec:indepOffLB}) was proposed by Kedad-Sidhoum et al.~\cite{HLP} as follows, where $C$ variables denote the completion time of tasks

\begin{align}
   \text{minimize } \quad & ~C_{LP} \notag \\ \text{subject to:} \quad
   & \sum_{j \in \mathcal{T}} \overline{p_j} x_j \leq m\cdot
   C_{LP}~~ \label{ci4}\\ & \sum_{j \in \mathcal{T}} \underline{p_j}
   (1-x_j) \leq k\cdot C_{LP}~~ \label{ci5}\\ & C_i + \overline{p_j}
   x_j + \underline{p_j} (1-x_j) \leq C_j && \forall j \in
   \mathcal{T}, i \in \Gamma^{-}(j) \label{ci1}\\ & \overline{p_j} x_j
   + \underline{p_j} (1-x_j) \leq C_j && \forall j \in \mathcal{T}:
   \Gamma^{-}(j)=\emptyset \label{ci2}\\ & C_j \leq C_{LP}~~ &&
   \forall j \in \mathcal{T} \label{ci3}\\ & x_j \in [0, 1] ~~ &&
   \forall j \in \mathcal{T} \label{ci6}\\ & C_j \geq 0~~~ && \forall
   j \in \mathcal{T} \notag
\end{align}

Let us denote this linear program by $LP_{prec}$.
Constraints (\ref{ci4}) and (\ref{ci5}) are the same as for $LP_{area}$, while Constraints (\ref{ci1}), (\ref{ci2}) and (\ref{ci3}) describe the critical path (CP) and ensure that precedence relation constraints between the tasks are satisfied.
As for $LP_{area}$, any schedule $S$ of makespan $C_{max}$ can be turned into a solution of $LP_{prec}$.
Thus, any optimal solution of this linear program provides a lower bound of the optimal makespan $C_{max}^*$.

\subsection{Lower Bounds in the On-line Setting}

\subsubsection{With independent tasks}
\begin{theorem}
\label{th:with-indep-tasks}
There is no on-line algorithm for scheduling independent tasks with a competitive ratio smaller than $2$.
\end{theorem}

This theorem was proposed by Chen et al.~\cite{Guochuan} and the proof is based on a simple instance with only two tasks.
Consider the special case with only one CPU and one GPU.
Suppose that the first task ready for scheduling has a processing time of 1 on both types of resources and, once it has been scheduled, a second task arrives, whose processing time on the resource where the first task is being processed is 1 and whose processing time on the other resource is 2. 
Then, for any decision of the scheduling algorithm for the second task the final makespan will be 2 while the optimal makespan is 1, proving the theorem.

\subsubsection{With precedence constraints}

\newcommand{\sqmkl}{\sqrt{m/k}} \newcommand{\sqmk}{\sqrt{\frac mk}}

Intuitively, the problem of on-line scheduling with precedence contraints on two types of processors is difficult.
Indeed, without knowing the successors of a task, how to decide
whether to accelerate this task on a GPU, or to save this rare
resource for a future task?
In a recent work, Canon et al.~\cite{europar18} confirmed this
intuition by proving that there is no constant-factor competitive
algorithm for this problem, as stated in Theorem~\ref{th:onlinedagLB}.
This bound completes the one from Theorem~\ref{th:with-indep-tasks},
which also holds with precedence constraints.

\begin{theorem}
\label{th:onlinedagLB}
There is no on-line algorithm for scheduling tasks with precedence constraints with a competitive ratio smaller than $\sqmkl$, for any value of $m$ and $k$.
\end{theorem}

The proof is based on the use of an adversary that builds a graph consisting of several rounds of $k\sqmkl$ independent tasks, whose processing time on CPU (resp. GPU) is $\sqmkl$ (resp. 1), assuming that $\sqmkl$ is an integer.
Any on-line algorithm therefore requires a time $\sqmkl$ to complete each round.
When a round is completed, the next one is revealed, where each new task is a successor of the last finishing task of the previous round.
After $r$ rounds, the processing time of any on-line algorithm is therefore at least $r\sqmkl$ whereas in comparison, an off-line algorithm, knowing in advance all precedence constraints, may allocate critical tasks on GPU and the others on CPU, thus achieving a makespan asymptotically close to $r$, by executing the CPU tasks of $\sqmkl$ rounds in parallel.

This theorem may not seem very robust.
Indeed, for instance, the optimal allocation is obvious as soon as the scheduler can detect terminal tasks, and in practice, such information may well be available to the scheduler, even if this violates the on-line setting assumptions.
However, \Cref{th:onlinedagLB} remains valid even if the upward rank of each task is known (see \Cref{sssec:HEFT}), except when $k=1$ in which case the lower bound is halved.

The idea behind the proof is to add a chain to the previous graph and to add a dependency from each task to this chain, so that all tasks have the same upward rank.
In the optimal schedule, this chain can be processed on an unused GPU without increasing the makespan.
Other generalizations of the bottom-level on a heterogeneous platform may exist, as each task has several possible computing times (see \Cref{sec:definitions}), but as all tasks are identical in our case, this result remains valid for any generalization of the bottom-level.

Even further, even if the processing times of all the descendants of each task are available to the scheduler, the authors prove there is no constant-factor competitive algorithm, and that any competitive ratio is lower bounded by $\Theta((m/k)^{1/4})$.
The authors also study the impact of an additional flexibility given to the scheduler: if the scheduler can kill a task and migrate it instantaneously to another node (such an operation is usually named spoliation), the results remain unchanged.
Even allowing unrealistic preemption and migration only halves the lower bounds.

\section{Approximation Algorithms and Heuristics for Independent Tasks}\label{sec:independent} 
 
We review here strategies proposed to schedule independent tasks onto a heterogeneous platform, both in off-line and on-line contexts.
 
\subsection{Off-line Results for Independent Tasks}\label{ssec:indepOFF} 
 
We start with the off-line setting, where all tasks are available at the beginning of the computation, and their characteristics (i.e., running times on both processor types) are known to the scheduler.
 
\subsubsection{DualHP}
\label{sssec:dualhp}

The \emph{DualHP} algorithm~\cite{fastbio}, presented in
\Cref{alg:dualHP}, uses the dual approximation technique presented in
\Cref{ssec:preliminaries} to obtain a guess ($\lambda$) on the optimal
makespan and then to build a schedule whose makespan is at most $2\lambda$.
This guess is used to solve a minimization knapsack problem that
assigns to GPUs tasks with a total load smaller than $(k+1) \lambda$ and
assigns all remaining tasks on CPUs.
At last, List Scheduling is applied on both sets of resources and
another iteration of the dual approximation technique is
performed.

At first, the initial step is a sorting which takes $O(n\log(n))$ operations.
Then, as stated in Section~\ref{sec:list-sched-appr}, the time
complexity of each application of List Scheduling is $O(n\log(m))$
and there are no more than
$\log_{2}(\frac{B_{max} - B_{min}}{\epsilon})$ iterations
(Section~\ref{sec:dual-approximation}).

\begin{proposition}
  If $\overline{W} \leq m \lambda$, there exists a feasible solution with a makespan at most $2\lambda$.
\end{proposition}

Since List Scheduling is used during the second step of the algorithm, the makespan on CPUs ($\overline{C_{max}}$) can be upper bounded as in Section~\ref{ssec:preliminaries}, $$ \overline{C_{max}} \leq \max_{j \in \mathcal{T}}
(\overline{p_j}x_j) + \frac{\sum_{j \in 
    \mathcal{T}}{(\overline{p_j}x_j)}}{m} $$ 
 
Then, since the processing time of all tasks assigned to a CPU is smaller than $\lambda$, then $\max_{j \in \mathcal{T}} (\overline{p_j}x_j) \leq \lambda$. Furthermore, our assumption ensures that $\sum_{j \in   \mathcal{T}}{(\overline{p_j}x_j)} \leq m \lambda$. Thus, $$ \overline{C_{max}} \leq 2 \lambda.$$
\smallskip 
 
A similar reasoning on the GPU side shows that $C_{max} \leq 2 \lambda$ and the following theorem holds.
 
\begin{theorem} 
  \emph{DualHP} (\Cref{alg:dualHP}) is a $2(1+\epsilon)$-approximation.
\end{theorem}

\begin{algorithm}
  \caption{DualHP~\cite{fastbio} \label{alg:dualHP}}
  \begin{algorithmic}[1]
    \STATE $L = \text{list of tasks sorted by non-increasing }\alpha_j=\frac{\overline{p_j}}{\underline{p_j}}$
    \STATE $\overline{W} \gets 0$
    \STATE $\underline{W} \gets 0$
    \FOR{each task $j \in L$}
      \IF{$\overline{p_j} > \lambda$}
        \IF{$\underline{p_j} > \lambda$}
          \RETURN "unfeasible guess".
        \ELSE
          \STATE $x_j \gets 0$
          \STATE $\underline{W} \gets \underline{W} + \underline{p_j}$
        \ENDIF
      \ELSE
        \IF{$\underline{p_j} > \lambda$}
          \STATE $x_j \gets 1$
          \STATE $\overline{W} \gets \overline{W} + \overline{p_j}$
        \ENDIF
      \ENDIF
    \ENDFOR
    \FOR{each remaining task $j \in L$}
      \IF{$\underline{W} < k \lambda$}
        \STATE $x_j \gets 0$
        \STATE $\underline{W} \gets \underline{W} + \underline{p_j}$
      \ELSE
        \STATE $x_j \gets 1$
        \STATE $\overline{W} \gets \overline{W} + \overline{p_j}$
      \ENDIF
    \ENDFOR
    \IF{$\overline{W} > m \lambda$ \textbf{or} $\underline{W} > (k+1) \lambda$}
      \RETURN "unfeasible guess".
    \ELSE
      \STATE Schedule all tasks using  List Scheduling with respect to the assignment variables $x_j$.
      \RETURN the current schedule.
    \ENDIF
  \end{algorithmic}
\end{algorithm}

\subsubsection{Two Families of Algorithms based on Dynamic Programming}
\label{sssec:DP}
 
Kedad-Sidhoum et al.~\cite{IJFCS2018} propose two families of algorithms that combine two techniques, namely dual approximation and dynamic programming.
The dual approximation ratios achieved by these algorithms are $\rho = \frac{2q+1}{2q} + \frac{1}{2qk}$ for $q>0$ and $\rho = \frac{2(q+1)}{2q+1}+ \frac{1}{(2q+1)k}$ for $q \geq 0$, which can be turned into $\rho(1+\epsilon)$-approximation algorithms using binary search, where $\epsilon$ is an arbitrarily small value corresponding to the threshold of the binary search.
The computational complexity of these algorithms is $O(n^2k^{q+1}m^q)$ and $O(n^2k^{q+2}m^{q+1})$, respectively.
Note that $q$ is a user-defined parameter, and larger values of $q$ lead to better accuracy at the expense of the complexity.
Moreover, algorithms from both families can be transformed into polynomial time approximation schemes by selecting for example $q=\frac{k+1}{2k\varepsilon}$ and $q=\frac{1}{2k\varepsilon}-1$, respectively.
 
\begin{figure}
  \centering \includegraphics[scale=.1]{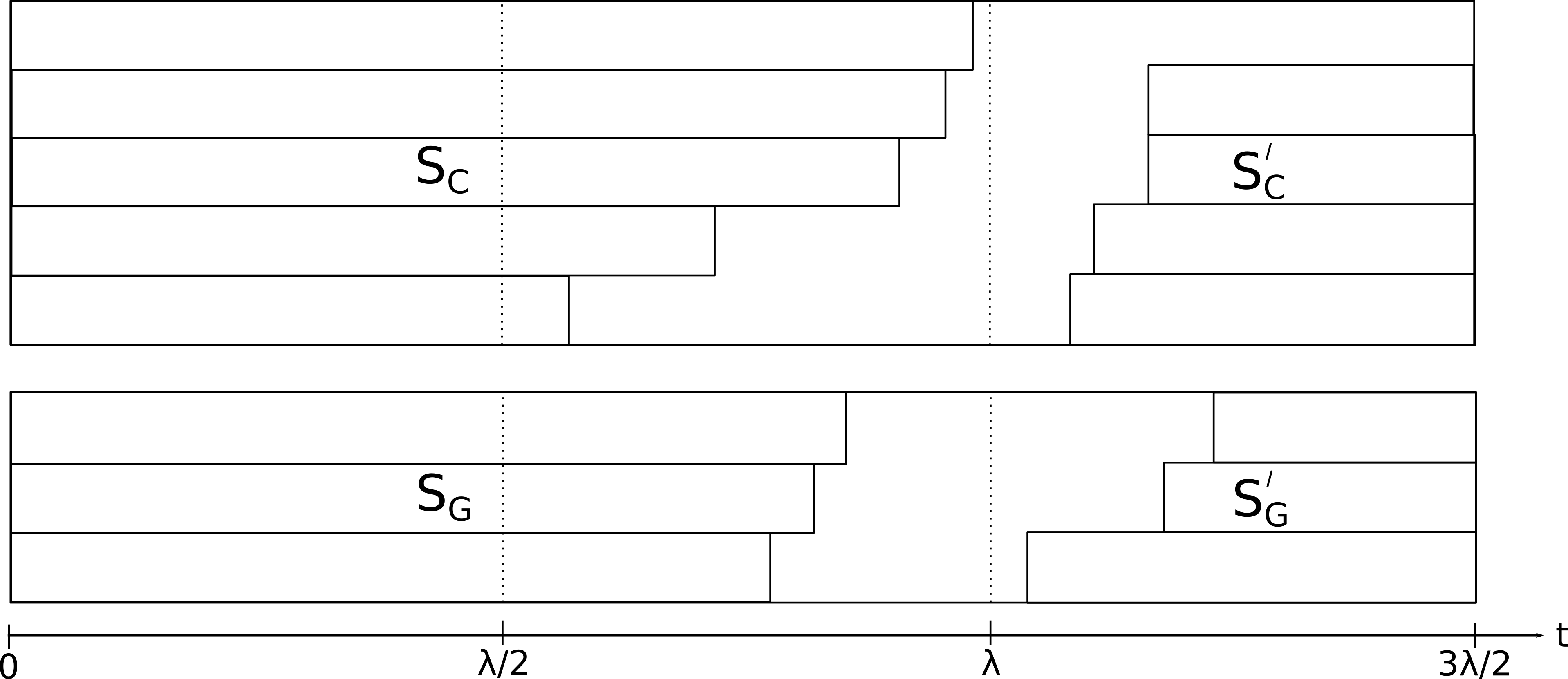} 
  \caption{Example of partitions of tasks into shelves for the 
    algorithms using dynamic programming (\Cref{sssec:DP}).} 
  \label{fig:shelves} 
\end{figure}

In the following, we explain the ideas behind these algorithms by briefly describing the algorithm \emph{DP$_\frac{3}{2}$} which corresponds to the first family with $q=1$ and a dual approximation ratio $\rho = \frac{3}{2} + \frac{1}{2k}$.
The idea of the algorithm is, given a guess $\lambda$ of the optimal makespan, to build a schedule whose makespan is at most $\frac{3\lambda}{2}$.
To achieve this result, tasks are partitioned into two shelves on the CPU side ($\mathcal{S}_C$ and $\mathcal{S}_C'$) and two shelves on the GPU side ($\mathcal{S}_G$ and $\mathcal{S}_G'$) as shown in \Cref{fig:shelves}.
Among the tasks assigned to a CPU, those whose processing time on a CPU is larger than $\frac{\lambda}{2}$ are placed in $\mathcal{S}_C$, and there cannot be more than $m$ of them in any solution.
The shorter tasks are placed in $\mathcal{S}_C'$, and the total execution time of all tasks that are assigned to CPU should be no more than $m\lambda$.
The shelves on the GPU are built similarly, with $\mathcal{S}_G$ containing tasks with $\underline{p_j} > \frac{\lambda}{2}$ (no more than $k$ of them) and $\mathcal{S}_G'$ all other tasks.

The problem of assigning the tasks on these four shelves can thus be formulated as a minimization multi-dimensional knapsack problem as follows (note that the objective function is strong since in practice we only need a feasible solution).
\begin{align*}
  \text{minimize } \quad & \sum_{j \in \mathcal{T}}{\overline{p_j}x_j}\\
  \text{subject to:} \quad & \sum_{j \in \mathcal{T}: \overline{p_j} > \frac{\lambda}{2}} {x_j} \leq m \\
  & \sum_{j \in \mathcal{T}}{\overline{p_j}x_j} \leq m \lambda \\
  & \sum_{j \in \mathcal{T}: \underline{p_j} > \frac{\lambda}{2}}{(1-x_j)} \leq k \\
  & \sum_{j \in \mathcal{T}}{\underline{p_j}(1-x_j)} \leq k \lambda \\
  & x_j \in \{0, 1\} \quad\forall j \in \mathcal{T}
\end{align*}

To solve the knapsack problem and obtain an assignment of the tasks to resources, a dynamic programming algorithm is used.
Before this, the time horizon on GPUs is discretized in time intervals of size $\frac{\lambda}{2n}$. Let $N$ be the number of these time intervals.

The dynamic programming formulation is then based on the parameters $j$ (the number of tasks already assigned), $\mu$ (the number of CPUs occupied in the shelf $S$), $\kappa$ (the number of GPUs occupied in shelf $G$) and $N$ (the number of busy time intervals on the GPUs).
This dynamic programming approach yields a polynomial time algorithm.
However, the knapsack problem can also be solved as an Integer Linear Program, which is not guaranteed to be solved in polynomial time but is much faster in practice.

The above idea can be generalized by partitioning the tasks into $q$ couples of shelves on each side.
For example, the shelf $\mathcal{S}_{C,h}$, $1 \leq h \leq q$, is composed of the tasks assigned to a CPU and such that $\frac{(2q-2)\lambda}{2q} \leq \overline{p_j} \leq \frac{(2q-1)\lambda}{2q}$, while the shelf $\mathcal{S}_{C,h}'$, $1 \leq h \leq q$, contains the tasks assigned to a CPU such that $\frac{(h-2)\lambda}{2q} \leq \overline{p_j} \leq \frac{h\lambda}{2q}$.
The shelves on the GPU side ($\mathcal{S}_{G,h},\mathcal{S}_{G,h}'$ for $1 \leq h  \leq q$) are defined in a similar way.
The problem of assigning the tasks into these generalized shelves can be also described  as a multi-dimensional minimization knapsack problem and it can be solved by a more complicated dynamic programming algorithm whose complexity depends on $q$ as described in the beginning of this section.

\subsubsection{HeteroPrio}
\label{sssec:heteroprio} 
 
The HeteroPrio scheduling algorithm was first introduced in a practical context for scheduling tasks in a Fast Multipole Method computation~\cite{BerengerHeteroPrio}, and showed good practical performance for this application in which precedence constraints are loose enough for most of tasks to be independent.
Later, Beaumont et al. provided a theoretical analysis of the algorithm~\cite{HeteroPrio} and proved approximation ratios when the tasks are independent.

The HeteroPrio algorithm is based on two main ideas.
The first idea, present in the original version~\cite{BerengerHeteroPrio}, consists in trying to get close to the area bound solution (presented above in \Cref{sssec:indepOffLB}) using a greedy scheduler.
This is done by sorting tasks by non-decreasing acceleration factors, and by having idle CPUs pick tasks from the beginning of the list, while idle GPUs are picking tasks from the end of the list.
 
The theoretical analysis shows that the partial solution produced by this first part of the algorithm is optimal in the following sense: 
 
\begin{lemma} 
  \label{lem.heteroprio}
  For any instance $I$, if all processors are busy up to time $t$, then $AreaBound(I) \geq t + AreaBound(I'(t))$, where $I'(t)$ is the sub-instance composed of parts of tasks not completed at time $t$.
\end{lemma} 
 
This lemma is the main ingredient of the approximation proof of HeteroPrio.
With the assumption that, for all tasks $j$, $\max(\overline{p_j}, \underline{p_j}) \leq C_{\max}^*$, this lemma alone allows one to prove a $2$-approximation ratio.
Indeed, if we denote by $t_{\text{FI}}$ the first idle time in the resulting schedule, Lemma~\ref{lem.heteroprio} shows that $t_{\text{FI}} \leq AreaBound$.
Furthermore, after time $t_{\text{FI}}$, each processor computes at most one task, and thus finishes not later than $t_{\text{FI}} + \max_j (\max(\overline{p_j}, \underline{p_j}))$, which concludes the proof.
 
However, this idea alone cannot provide an approximation guarantee in the general case.
Indeed, this strategy produces a list schedule, where no processor is left idle if a task is available, and we have proved in \Cref{sssec:indepOffLB} that List Scheduling can lead to arbitrarily bad results in this heterogeneous setting.
 
The second idea to provide an approximation guarantee, introduced by Beaumont et al.~\cite{HeteroPrio}, is \emph{spoliation}, which enables re-scheduling a task from a resource to an idle resource, in order to complete it sooner.
They introduce a simple greedy scheme where the first idle resource spoliates tasks scheduled to finish last, and prove that such a greedy spoliation strategy is enough to overcome the bad performance of List Scheduling, and to obtain a constant factor approximation ratio.
\Cref{alg:HeteroPrio} presents the pseudo-code of the complete version of HeteroPrio.
 
\begin{algorithm} 
  \caption{HeteroPrio~\cite{BerengerHeteroPrio,HeteroPrio} \label{alg:HeteroPrio}} 
  \begin{algorithmic}[1] 
    \STATE $L \gets \text{list of tasks sorted by non-decreasing }\alpha_j=\frac{\overline{p_j}}{\underline{p_j}}$ 
    \WHILE{not all tasks are completed} 
    \STATE $t \gets$ first time a resource $i$ is idle 
    \IF{$L$ is non empty} 
      \IF{idle resource $i$ is a CPU} 
        \STATE Pop task $j$ from the head of $L$.
      \ELSE 
        \STATE Pop task $j$ from the tail of $L$.
      \ENDIF 
      \STATE Start task $j$ on resource $i$ at time $t$.
    \ELSE 
      \STATE $S \gets \{\text{tasks assigned to other resources, that would finish earlier if started on $i$ at time $t$}\}$ 
      \IF{$S$ is non empty} 
        \STATE $j \gets $  task from $S$ with highest finish time 
        \STATE Unassign $j$ and start it on resource $i$ at time $t$.
      \ELSE 
        \RETURN the current schedule.
      \ENDIF 
      \ENDIF 
      \ENDWHILE 
  \end{algorithmic} 
\end{algorithm}

\begin{figure}
  \centering \subfloat[Optimal schedule]{ 
    \begin{tikzpicture}[scale=1, every node/.style={transform shape}] 
      \tikzstyle{legend}=[below=0.4, anchor=mid] 
      \node at (0,0) [anchor=south west, draw,minimum width=3cm, 
        minimum height=1cm,fill=white, text width=2.75cm, text badly 
        centered] (t2) {Optimal schedule for $T_2$}; 
      \node at (0, 1.51cm) [anchor=south west, draw,minimum width=1.99cm, 
        minimum height=2.5cm,fill=white] (t3) {Set $T_3$}; 
      \node[right=-0.01cm of t3, draw, minimum width=1cm, minimum 
        height=2.5cm,fill=white] (t4) {$T_4$}; 
      \foreach \i in {0, 0.2, ..., 0.8} 
        \node at (0, 4+\i)[anchor=south west, draw, fill=white,minimum 
          width=3cm, minimum height=0.2cm](t1) {}; 
      \node at (1.5,4.5) [fill=white] {$T_1$}; 
      \node at (-0.3, 3.25) [anchor=south, rotate=90]{$m=k^2$ CPUs};
      \node at (-0.3, 0.5)[anchor=south, rotate=90] {$k$ GPUs};
      \draw[->] (-0.1, -0.1) -- (0, -0.1) -- 
        node[pos=0, legend] {$0$} 
        node [pos=0.86,legend] {$k$} (3.5, -0.1) 
        node[right] {$t$}; 
      \foreach \x in {0, 3} \draw (\x, -0.1) -- (\x, -0.2); 
      \path (-0.1, -0.8) -- (2, -0.8); 
    \end{tikzpicture} 
  } 
  \subfloat[HeteroPrio when the last task can be spoliated]{
    \begin{tikzpicture}[decoration={brace, amplitude=4}, scale=1, every node/.style={transform shape}] 
      \tikzstyle{legend}=[below=0.4, anchor=mid] 
      \node at (0,0) [anchor=south west, draw,minimum width=1.8cm, 
        minimum height=1cm,fill=white] (t4) {Set $T_4$}; 
      \node at (0,1.5cm) [anchor=south west, draw, minimum 
        width=1.8cm, minimum height=3.5cm,fill=white] (t3) {Set $T_3$}; 
      \foreach \i in {0, 0.2, ..., 0.8} 
        \node at (1.8, 1.5+\i) [anchor=south west, 
          draw=gray, fill=white, minimum width=7cm, 
          minimum height=0.2cm](t2) {}; 
      \node at (5.3,2) [fill=white, anchor=center] {$T_2$ (aborted)}; 
      \draw[decorate, line width=0.5pt] (1.8, 2.6) -- 
        node[above=0.3cm]{$\frac{rk}{3}$} (8.8, 2.6); 
      \foreach \i in {0, 0.2, ..., 0.8} 
        \node at (1.8, \i)[anchor=south west, draw, minimum 
          width=1cm,minimum height=0.2cm, inner sep=0pt](t1) {}; 
      \node at (2.3, 0.5) [fill=white] {$T_1$}; 
      \node at (2.8, 0) [anchor=south west, draw,minimum height=1cm, 
        minimum width=3cm] {Bad $T_2$ schedule}; 
      \node at (5.8, 0) [anchor=south west, draw, minimum 
        height=0.2cm, minimum width=3cm, inner sep=0pt] {};      
      \draw[->] (-0.1, -0.1) -- (0, -0.1) -- 
      node[legend, pos=0] {$0$} 
      node[legend, pos=0.2] {$x$} 
      node[legend, pos=0.31] {$x+\frac{k}{r}$} 
      node[legend, pos=0.64]{$x+\frac{k}{r}+k-1$} 
      node[legend, pos=0.95]{$x+\frac{k}{r}+2k-1$} 
      (9.3, -0.1)  node[right] {$t$}; 
      \foreach \x in {0, 1.8, 2.8, 5.8, 8.8} 
        \draw (\x, -0.1) -- (\x, -0.2); 
 
      \draw[dashed, line width=.3] (0.01,0.005) rectangle (8.81,1.01);
      \draw[dashed, line width=.3] (0.01,1.505) rectangle (8.81,5.01);
    \end{tikzpicture} 
  }
  \caption{Optimal and HeteroPrio on the worst case instance (\Cref{sssec:heteroprio}).} 
    \label{fig:heteroprio.worst.case} 
\end{figure}
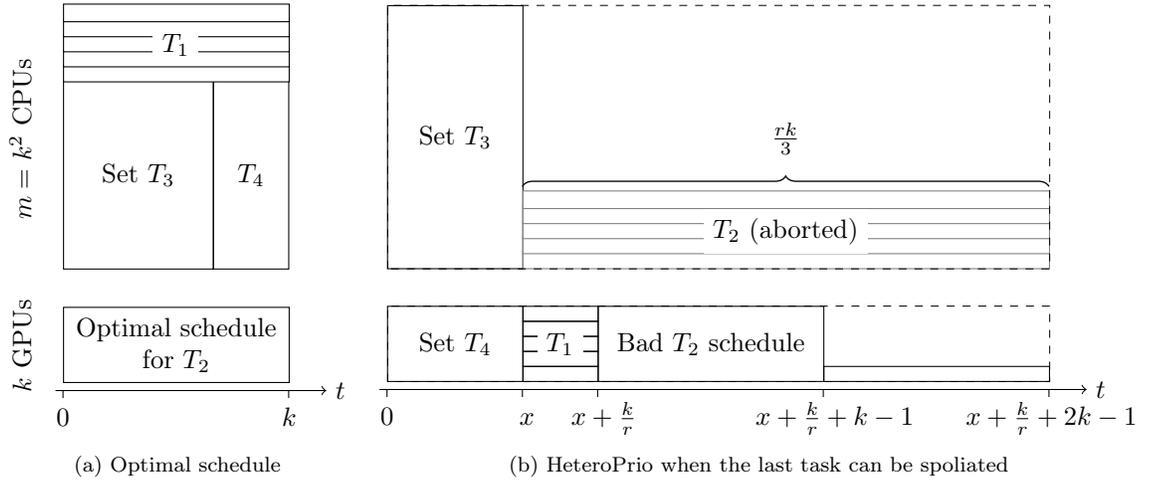 
 
To explain the underlying idea of the approximation proof of HeteroPrio, let us first present the worst-case example proposed by the authors and depicted in Figure~\ref{fig:heteroprio.worst.case}, where $x/k$ tends to $1$, and $r$ tends to $3+2\sqrt{3}$ when $k$ approaches $+\infty$).
The instance consists of four sets of tasks.
Tasks in the set $T_3$ have an acceleration factor of $1$, and tasks in the sets $T_1$ and $T_4$ have an acceleration of factor $r > 1$.
Tasks in $T_1$ exhibit long execution times, and tasks in $T_3$ and $T_4$ are small enough so that we can concentrate on their total load only.
Tasks in $T_2$ have specially crafted execution times on GPUs, so that there exists a list schedule for these tasks on GPUs with makespan $2-\frac{1}{k}$ times larger than the optimal, and set $T_2$ does not contain small tasks.
Tasks of $T_2$ all have the same execution time on CPU; their acceleration factors differ, but all are between $1$ and $r$.
The actual execution times of tasks in the instance are chosen so that the schedule depicted on the left part of Figure~\ref{fig:heteroprio.worst.case} is feasible with makespan $C_{max}^*=k$.
On the other hand, a possible schedule for HeteroPrio is depicted on the right part of this figure: tasks of sets $T_3$ and $T_4$ are executed first (this takes almost a time $k$), then tasks of $T_1$ start on the GPUs and tasks of $T_2$ start on the CPUs.
These tasks from $T_2$ have very long execution times on CPUs, so they get spoliated by the GPUs once GPUs are done with $T_1$ tasks.
This results in a list schedule on the GPUs for tasks from $T_2$, which can take up to $2k$ units of time.

To make this instance feasible, the largest possible value for the execution time of tasks in $T_1$ on GPUs is given by $(\frac{2}{\sqrt{3}} - 1)k$.
This implies that the approximation ratio of HeteroPrio is at least $1 + (\frac{2}{\sqrt{3}} - 1) + 2$, where the first part comes from the work done before the first idle time, the second part comes from executing the remaining tasks on the GPUs given their maximum acceleration factor, and the last part comes from List Scheduling bounds.
More details on this proof can be found in the original paper~\cite{HeteroPrio}.\\
 
The proof for the constant approximation ratio of HeteroPrio follows the same structure: Lemma~\ref{lem.heteroprio} provides a bound for the time before some resource becomes idle.
The length of tasks remaining on the GPUs is bounded, and since spoliated tasks are too long to be executed on CPUs in any optimal schedule, the fact that HeteroPrio is a List Scheduler for them ensures that it takes at most twice the optimal makespan.
However the bound on the length of the tasks remaining on the GPUs after the first idle time is less than in the above counter-example ($\sqrt{2} - 1$ instead of $\frac{2}{\sqrt{3}} - 1$).
This comes from the fact that in their proof~\cite{HeteroPrio}, the authors have not been able to take the amount of work left after the first idle time into account in the equations that express the constraint over the acceleration factors of the tasks remaining on the GPUs.

Overall, the following theorem summarizes HeteroPrio's approximation ratios in the different cases.
 
\begin{theorem} 
  \label{thm:heteroprio} 
  The approximation ratio of HeteroPrio (\Cref{alg:HeteroPrio}) is at least $2 + 
  \frac{2}{\sqrt{3}} \sim 3.15$ and at most $2 + \sqrt{2} \sim 
  3.41$. In the special case when  $\overline{p_j} \leq C_{max}^*$ and 
  $\underline{p_j} \leq C_{max}^*$ for any task $j \in \mathcal{T}$, 
  HeteroPrio is a 2-approximation algorithm. 
\end{theorem}

HeteroPrio has a time complexity of $O(n \log(n) + n\log(m) +
m\log(m))$, where the first term is for sorting the set of tasks at
the beginning, the second is for retrieving the processor with
smallest completion time and the last term is for scheduling the
spoliated tasks: since $m\geq k$, there are at most $m$ candidate
tasks (spoliation only occurs from one resource type to the other),
and sorting these candidate tasks in an appropriate data structure
leads to a $log(m)$ amortized cost for finding the best candidate.

\subsubsection{BalancedEstimate and BalancedMakespan}
\label{sssec:balanced} 
\newcommand{\jmax}[1]{\textnormal{jmax}(#1)} 
 
The objective of BalancedEstimate and BalancedMakespan, proposed by Canon et al.~\cite{BalancedEuropar}, is very similar to the one of HeteroPrio.
However, these heuristics are based on slightly different ideas to construct the schedule.
Both heuristics follow the same principle, which consists in two main steps.
For the sake of clarity, we first explain the BalancedEstimate heuristic, and then point out the differences with BalancedMakespan.
First, the allocation $x_j$ of each task $j \in \mathcal{T}$ is computed to decide whether a task should be computed on a CPU or on a GPU.
Then, a precise schedule of the tasks allocated to each resource type is computed.
 
The first step is the most critical.
In this phase, described in Algorithm~\ref{algo.balancedalloc}, BalancedEstimate starts from the allocation where each task is put on its favorite resource type, i.e., the processor type on which it has the minimum processing time.
Then, this allocation is refined to balance the load of the two processor types.
Without loss of generality, we assume that GPUs are more loaded than CPUs (otherwise we exchange the role of CPUs and GPUs).
To rebalance the load, we consider each task allocated on GPUs, starting from the one which suffers the least from being on its non-favorite resource, that is, the task $j$ such that $\alpha_j$ is minimal.
Each of these tasks is iteratively moved to CPUs, and two special allocations are remembered:
\begin{itemize} 
\item The allocation $x^{\mathit{best}}$ leading to the best \emph{estimated} makespan $\mathit{Est}(\mu)$ (allocation cost estimate, defined below); 
\item The allocation $x^{\mathit{inv}}$ obtained where CPUs become overloaded.
\end{itemize} 
During this iterative process, we also take care of special tasks that would significantly degrade the makespan if moved to CPUs: when one of the tasks moved to CPU dominates the makespan (i.e., when the processing time on CPU of that task is greater than or equal to the estimated makespan), it is moved back to GPUs.

Initially, tasks are allocated to their favorite resource type (Lines~\ref{line:initbegin}--\ref{line:initend} of Algorithm~\ref{algo.balancedalloc}).
GPUs are assumed to have the largest average load, otherwise processor types are swapped (Line~\ref{line:reverse}).
Tasks are sorted by non-decreasing acceleration ratio (Line~\ref{line:sort}), such that the first task to move to CPUs is $j_{\textnormal{start}}$, as defined on Line~\ref{line:istart}.
 
In order to define the allocation cost estimate, we first extend the notation $\overline{W}$ and $\underline{W}$ so that $\overline{W}(x)$ (resp. $\underline{W}(x)$) denotes the total overall workload on all CPUs (resp. GPUs) for the allocation $x$.
We also define the maximum processing time $\overline{M}(x)$ (resp. $\underline{M}(x)$) of tasks allocated on CPUs (resp. GPUs) as follows:
$$ 
\overline{M}(x)=\max_j x_j \overline{p_j} \quad\textnormal{and}\quad \underline{M}(x)=\max_j (1-x_j) \underline{p_j}.
$$ 
BalancedEstimate relies on the maximum of the four previous quantities presented above to estimate the makespan of an allocation.
More precisely, the \emph{allocation cost estimate} is defined as follows:
$$\mathit{Est}(x)=\max\left(\frac{\overline{W}(x)}{m},\frac{\underline{W}(x)}{k},\overline{M}(x),\underline{M}(x)\right).$$
This estimation of the makespan is used to define the best allocation seen so far, denoted $x^{\mathit{best}}$ and updated in Line~\ref{line:elitism} of Algorithm~\ref{algo.balancedalloc}.
Line~\ref{line:inversion} defines the allocation $x^{\mathit{inv}}$ leading to an inversion of the largest load, while Line~\ref{line:move} moves the current task from GPUs to CPUs.
$\jmax{x}$ denotes the index of the largest task allocated to a CPU but that would be more efficient on a GPU:
$$\jmax{x}=\argmax_{j: x_j=1 \textnormal{~and~} \alpha_j>1} \overline{p_j}.
$$
Finally, a \emph{dominating} task $j$ verifies $j=\jmax{x}$ and $\mathit{Est}(x)=\overline{p_{\jmax{\mu}}}$.
Line~\ref{line:dominating} of Algorithm~\ref{algo.balancedalloc} checks if there exists a dominating task and, if any, moves it back to GPUs.

 \begin{algorithm}[tb] 
 \caption{Balanced Allocation used in BalancedEstimate} 
 \label{algo.balancedalloc} 
  \begin{algorithmic}[1] 
\FOR{$j=1\ldots n$}\label{line:initbegin} 
\IF{$\alpha_j=\frac{\overline{p_j}}{\underline{p_j}} <1$}
    \STATE $x_j\gets 1$ 
  \ELSE 
    \STATE $x_j\gets 0$ 
  \ENDIF 
\ENDFOR\label{line:initend} 
\IF{$\frac{\overline{W}(x)}{m} > \frac{\underline{W}(x)}{k}$} 
  \STATE {Switch processor types.\label{line:reverse}} 
\ENDIF 
\STATE $x^{\mathit{best}}\gets x$ 
\STATE Sort tasks by non-decreasing $\alpha_j.$\label{line:sort}
\STATE $j_{\textnormal{start}} \gets \min\{j: x_j=0\}$ \label{line:istart}  
\FOR{$j=j_{\textnormal{start}}\ldots n$} 
  \IF{$\frac{\overline{W}(x)}{m} \leq \frac{\underline{W}(x)}{k}$ \textnormal{\textnormal{\textbf{and}}} $\frac{\overline{W}(x)+\overline{p_j}}{m} > \frac{\underline{W}(x)-\underline{p_j}}{k}$} 
    \STATE $x^{\mathit{inv}}\gets x$ \label{line:inversion} 
  \ENDIF 
  \STATE $x_j \gets 1$\label{line:move}  
  \IF{$\mathit{Est}(x)<\mathit{Est}(x^{\mathit{best}})$\label{line:elitism}} 
    \STATE $x^{\mathit{best}}\gets x$  
  \ENDIF 
  \IF{$\mathit{Est}(x) =\overline{p_{\jmax{x}}}$\label{line:dominating}} 
    \STATE $x_{\jmax{x}}\gets 0$ \label{line:back} 
  \ENDIF 
\ENDFOR 
\IF{$x^{\mathit{inv}}$ is not defined} 
\STATE $x^{\mathit{inv}}\gets x$ 
\ENDIF 
\RETURN $(x^{\mathit{best}},x^{\mathit{inv}})$ 
\end{algorithmic} 
\end{algorithm}

\begin{algorithm}[tb] 
\caption{BalancedEstimate~\cite{BalancedEuropar}} 
\label{algo.balancedestimate} 
\begin{algorithmic}[1] 
\STATE Compute $(x^{\mathit{best}}, x^{\mathit{inv}})$ using Algorithm~\ref{algo.balancedalloc}.
\FOR{allocation $x$ in $(x^{\mathit{best}}, x^{\mathit{inv}})$}
  \STATE Schedule tasks $\{j:x_j=1\}$ on CPUs using LPT.
  \STATE Schedule tasks $\{j:x_j=0\}$ on GPUs using LPT.
\ENDFOR 
\RETURN the schedule that minimizes the global makespan.
\end{algorithmic} 
\end{algorithm} 
 
\begin{figure}[t] 
  \centering 
  \subfloat[Optimal schedule]{
  \begin{tikzpicture}[ 
    node distance=0, 
    every node/.style={inner sep=0pt,thick,minimum height=.5cm,outer sep=0pt}, 
    long/.style={rectangle,minimum width=2.5cm,draw=black}, 
    longlong/.style={rectangle,minimum width=3cm,draw=black}, 
    short/.style={rectangle,minimum width=.5cm,draw=black}, 
    shorteps/.style={rectangle,minimum width=.8cm,draw=black}, 
    ]
    \tikzstyle{legend}=[below=0.4, anchor=mid]
    \node (t3) [longlong] {$m$};
    \node (t1) [long, above=.5cm of t3.north west,anchor=south west] {$m-1$};
    \node (t1s) [short,right=of t1] {$1$};
    \node (dots1) [above=of t1] {\textbf{.}};
    \node (dots2) [above=of dots1] {\textbf{.}};
    \node (dots3) [above=of dots2] {\textbf{.}};
    \node (t2) [long,above=of dots3] {$m-1$};
    \node (t2s) [short,right=of t2] {$1$};

    \draw[dashed, line width=.6] (-1.5,0.75) rectangle (1.5,3.25);

    \node at (-1.8, 2) [anchor=south, rotate=90]{$m$ CPUs};
    \node at (-1.8, 0)[anchor=south, rotate=90] {$1$ GPU};

    \draw[->,thick] ($ (t3.south west) + (0,-.3cm) $) --
      node[legend, pos=0] {$0$}
      node[legend, pos=0.9] {$m$}
      ($ (t3.south east) + (0.3,-.3cm) $) node[right=.1] {$t$};
    \draw ($ (t3.south west) + (0,-.4cm) $) --
      ($ (t3.south west) + (0,-.2cm) $);
    \draw ($ (t3.south east) + (0,-.4cm) $) --
      ($ (t3.south east) + (0,-.2cm) $);

  \end{tikzpicture}
  }\hspace{4em}
  \subfloat[Schedule built for $x^{\mathit{best}}=x^{\mathit{inv}}$]{
  \begin{tikzpicture}[ 
    node distance=0, 
    every node/.style={inner sep=0pt,thick,minimum height=.5cm,outer sep=0pt}, 
    long/.style={rectangle,minimum width=2.5cm,draw=black}, 
    longlong/.style={rectangle,minimum width=3cm,draw=black}, 
    short/.style={rectangle,minimum width=.5cm,draw=black}, 
    shorteps/.style={rectangle,minimum width=.8cm,draw=black}, 
    ]
    \tikzstyle{legend}=[below=0.4, anchor=mid]
    \node (gs1) [shorteps] {$1+\epsilon$}; 
    \node (gs2) [shorteps,right=2.5cm of gs1.west, anchor=east] {$1+\epsilon$}; 
    \draw[draw=none] (gs1.east) -- (gs2.west) node[pos=.25] {\textbf{.}} 
    node[pos=.5] {\textbf{.}} node[pos=.75] {\textbf{.}}; 

    \node (g1) [long,above=.5cm of gs1.north west, anchor=south west] {$m-1$}; 
    \node (gdots1) [above=of g1] {\textbf{.}};
    \node (gdots2) [above=of gdots1] {\textbf{.}};
    \node (g1bis) [long,above=of gdots2] {$m-1$};
    \node (gs1bis) [short,right= of g1bis] {$1$};
    \node (g2) [long,above=of g1bis] {$m-1$};
    \node (g3) [long,right=of g2] {$m-1$};

    \draw[dashed, line width=.6] (-.4,0.75) rectangle (4.6,3.25);
    \draw[dashed, line width=.6] (-.4,-.25) rectangle (4.6,.25);

    \draw[->,thick] ($ (gs1.south west) + (0,-.3cm) $) --
      node[legend, pos=0] {$0$}
      node[legend, pos=.93] {$2m-2$}
      node[legend, pos=.465] {$m-1$}
    ($ (g3.east |- gs1.south) + (0.3,-.3cm) $) node[right=.1] {$t$};

    \draw ($ (gs1.south west) + (0,-.4cm) $) --
      ($ (gs1.south west) + (0,-.2cm) $);
    \draw ($ (g3.east |- gs1.south) + (0,-.4cm) $) --
      ($ (g3.east |- gs1.south) + (0,-.2cm) $);
    \draw ($ (gs2.south east) + (0,-.4cm) $) --
      ($ (gs2.south east) + (0,-.2cm) $);
  \end{tikzpicture}
  }
  \caption{Tightness of BalancedEstimate, achieved with $m>1$ CPUs,
    $k=1$ GPU and two types of tasks: $m$ tasks with costs
    $\overline{p_j}=1$ and $\underline{p_j}=1+\epsilon$
    (with $\epsilon < \frac{1}{m-1}$), and $m+1$ tasks with
    costs $\overline{p_j}=m-1$ and $\underline{p_j}=m$.
    After switching processor types, BalancedEstimate builds a
    schedule with makespan $2m-2$, whereas the optimal is $m$ (\Cref{sssec:balanced}).
  }
  \label{fig:tightness-balancedEstimate} 
\end{figure}
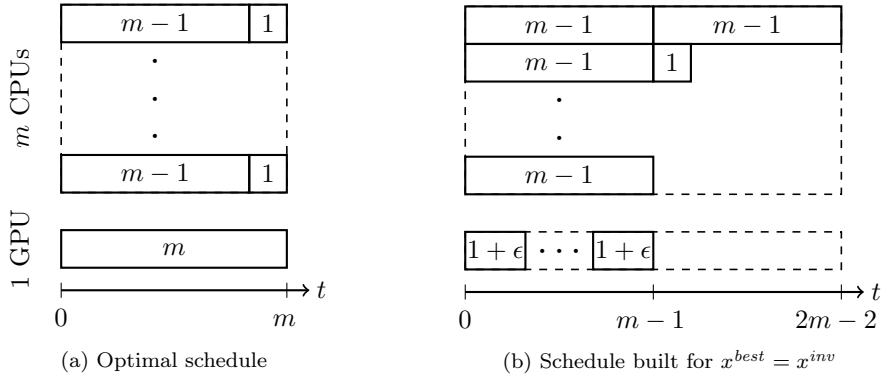

The scheduling phase (Algorithm~\ref{algo.balancedestimate}) computes, for each resource type, an LPT schedule for both $x^{\mathit{best}}$ and $x^{\mathit{inv}}$.
The final result is the schedule with the minimum makespan.
This results in a 2-approximation algorithm, as stated below.
Figure~\ref{fig:tightness-balancedEstimate} presents an example which shows the tightness of the approximation ratio.
 
\begin{theorem} 
  BalancedEstimate (Algorithm~\ref{algo.balancedestimate}) is a 2-approximation, and this ratio is tight.
\end{theorem} 
 
The most costly operation of \Cref{algo.balancedalloc} is the computation of the allocation cost estimate (Line~\ref{line:elitism}), which is in $O(n\log(n))$.
The time complexity of \Cref{algo.balancedestimate} is $O(n(\log(m)+\log(k)))$, which makes the overall complexity of BalancedEstimate $O(n\log(nm))$.
 
BalancedMakespan, described in Algorithm~\ref{algo.balancedmakespan} is a slightly more costly variant of BalancedEstimate: instead of using the allocation cost estimate during the allocation phase, it simulates the LPT policy and computes the exact resulting makespan. It has the same approximation ratio, but a larger time complexity $O(n^2\log(nm))$ and performs better in practice.
In Algorithm~\ref{algo.balancedmakespan}, the makespan of the schedule obtained using LPT for both CPUs and GPUs on allocation $x$ is denoted by $LPT(x)$.
 
\begin{algorithm}[tb] 
\caption{BalancedMakespan~\cite{BalancedEuropar}} 
\label{algo.balancedmakespan} 
\begin{algorithmic}[1] 
\FOR{$j=1\ldots n$} 
\IF{$\alpha_j =\frac{\overline{p_j}}{\underline{p_j}} <1$}
    \STATE $x_j\gets 1$ 
  \ELSE 
    \STATE $x_j\gets 0$ 
  \ENDIF 
\ENDFOR 
\IF{$\frac{\overline{W}(x)}{m} > \frac{\underline{W}(x)}{k}$} 
  \STATE Switch processor types. 
\ENDIF 
\STATE $x^{\mathit{best}}\gets x$ 
\STATE Sort tasks by non-decreasing $\alpha_j$.
\STATE $j_{\textnormal{start}} \gets \min\{j: x_j=0\}$  
\FOR{$j=j_{\textnormal{start}}\ldots n$} 
  \STATE  $x_j \gets1$  
  \IF{$LPT(x)<LPT(x^{\mathit{best}})$} 
    \STATE $x^{\mathit{best}}\gets x$  
  \ENDIF 
  \IF{$\mathit{Est}(x) =\overline{p_{\jmax{\mu}}}$} 
    \STATE $x_{\jmax{x}}\gets 0$  
  \ENDIF 
  \IF{$LPT(x)<LPT(x^{\mathit{best}})$} 
    \STATE $x^{\mathit{best}}\gets x$  
  \ENDIF 
\ENDFOR 
\RETURN the schedule produced using LPT for both CPUs and GPUs on allocation $x^{\mathit{best}}$.
\end{algorithmic} 
\end{algorithm}

\subsubsection{CLB2C}
\label{sssec:clb2c} 
 
\emph{Centralized Load Balancing for Two Clusters}, CLB2C in short, is a low-complexity scheduling heuristic proposed by Cheriere and Saule~\cite{CLB2C}.
The algorithm first sorts tasks by increasing acceleration factor.
It then compares the allocation of the first task on the soonest available CPU to the allocation of the last task on the soonest available GPU.
The choice leading to the smallest increase in makespan among both machines is chosen in the final schedule and the task is removed from the list.
The process continues until there is no more task to schedule.
 
\Cref{algo.CLBtwoC} details the steps of CLB2C, where $C(i)$ denotes the time where all jobs of processor $i$ have completed.\\ 
 
\begin{algorithm}[htbp] 
\caption{Centralized Load Balancing for Two Clusters (CLB2C)~\cite{CLB2C}} 
\label{algo.CLBtwoC} 
\begin{algorithmic}[1] 
\STATE Sort tasks by non-decreasing $\alpha_j=\frac{\overline{p_j}}{\underline{p_j}}$. 
\STATE $j^{\min}\gets 1$ 
\STATE $j^{\max}\gets n$ 
\WHILE{$j^{\min}\le j^{\max}$} 
\STATE Select the CPU $i^{\mathit{cpu}}$ such that $C(i^{\mathit{cpu}})=\min_{i\in CPU} C(i)$. 
\STATE Select the GPU $i^{\mathit{gpu}}$ such that $C(i^{\mathit{gpu}})=\min_{i\in GPU} C(i)$. 
\IF{$C(i^{\mathit{cpu}})+\overline{p_{j^{\min}}}\le C(i^{\mathit{gpu}})+\underline{p_{j^{\max}}}$} 
\STATE Allocate task $j^{\min}$ on machine $i^{\mathit{cpu}}$. 
\STATE $j^{\min}\gets j^{\min}+1$ 
\ELSE 
\STATE Allocate task $j^{\max}$ on machine $i^{\mathit{gpu}}$. 
\STATE $j^{\max}\gets j^{\max}-1$ 
\ENDIF 
\ENDWHILE 
\end{algorithmic} 
\end{algorithm} 
 
In the original publication, the authors prove that CLB2C is a 2-approximation algorithm when no task has a processing time larger than the optimal makespan.
The authors argue that this is a rather natural assumption for distributed scheduling, when there is a large number of tasks to schedule.
 
\begin{theorem} 
  CLB2C (\Cref{algo.CLBtwoC}) is a 2-approximation algorithm provided that 
  $\overline{p_j} \leq C_{max}^*$ and 
  $\underline{p_j} \leq C_{max}^*$ for any task $j \in \mathcal{T}$.
\end{theorem} 
 
The proof of this approximation ratio relies on identifying the earliest completion time $t$ at the instant the last task is scheduled.
The authors notice that all processors are busy in time interval $[0,t]$ and that the load balancing is optimal in this interval (moving a task can only increase the overall work), such that $t\leq \mathit{OPT}$.
Then, they manage to upper bound the extra time between $t$ and the final completion time by an extra $\mathit{OPT}$.
 
With an appropriate data structure such as a binary heap to retrieve
the CPU (resp.
GPU) with smallest completion time and update it in time $O(\log m)$
(resp.
$O(\log k)$), the total complexity of CLB2C is
$O(n (\log (nm)))$.

\subsection{On-line Setting for Independent Tasks}\label{ssec:indepON} 
 
We now move to the on-line setting: tasks are still independent, however they are submitted over time.
Recall that a task is scheduled immediately and irrevocably upon its arrival.
 
\subsubsection{PG, LG and MG}
\label{sssec:LGMG} 
 
In 2003, Imreh~\cite{Imreh2003} proposed two simple heuristics and a $4-2/m$-competitive algorithm for this problem.
 
The first heuristic, named \emph{Post Greedy} (PG in short) and presented in \Cref{alg:PG}, schedules each task on the machine where it will be finished earliest.
The time complexity of this operation is $O(\log(m))$.
This intuitive idea has often been used and it is named \emph{Earliest Finish Time} (EFT) or \emph{Earliest Completion Time} (ECT)~\cite{leung2004handbook}.
 
\begin{theorem} 
The competitive ratio of PG (\Cref{alg:PG}) is at least $\lfloor \frac{m}{k} \rfloor$.
\end{theorem} 
The proof is as follows: Let $\varepsilon>0$ be as small as desired, and consider a sequence of tasks decomposed into $\lfloor m/k\rfloor$ rounds.
Each round consists of $k$ tasks of processing times $\overline{p_j}=1+\varepsilon$ and $\underline{p_j}=1$, followed by $m$ tasks of processing times $\overline{p_j}=1$ and $\underline{p_j}=\varepsilon$.
PG schedules the first part of each round on GPUs and the second part on CPUs, achieving a makespan equal to the number of rounds, i.e., $\lfloor m/k\rfloor$.
It is however possible to achieve a makespan of 1 by scheduling each task on the opposite resource type.
The problem with this heuristic is that a task may be scheduled on one of the rare GPUs even if its processing time is only slightly reduced.
Then, tasks that could be significantly accelerated on a GPU end up
scheduled on an idle CPU if all GPUs are busy.
Both GPU and CPU are therefore not efficiently used.

\begin{algorithm} 
\caption{Post Greedy (PG)~\cite{Imreh2003} or Earliest Completion Time (ECT)~\cite{leung2004handbook}}\label{alg:PG} 
\begin{algorithmic}[1] 
\STATE Upon arrival of task $j$: 
\STATE Schedule task $j$ on the machine where it will be finished the earliest.
\end{algorithmic} 
\end{algorithm}

The second heuristic, named \emph{Load Greedy} (LG) and presented in \Cref{alg:LG}, assigns each task to the resource type on which the ratio of its processing time divided by the number of processors on that resource type is the smallest.
With this method, tasks are assigned to a GPU only if they are accelerated enough, so that the throughput of GPUs is higher for many identical tasks.
However, it is obvious that the competitive ratio is larger than $m/k$: a single task with $\overline{p_j}=m/k$ and $\underline{p_j}=1+\varepsilon$ will be scheduled on CPU.
Imreh proves that the competitive ratio is actually equal to $2+\frac{m-1}{k}$.
The most costly operation consists in finding the earliest available
idle resource on Line~\ref{line:LG-sched}, which time complexity is
$O(\log(m))$.
 
\begin{algorithm} 
\caption{Load Greedy (LG)~\cite{Imreh2003}}\label{alg:LG} 
\begin{algorithmic}[1] 
\STATE Upon arrival of task $j$: 
\IF {${\overline{p_j}}/{m} \geq {\underline{p_j}}/{k}$} 
\STATE $x_j\gets0$ 
\ELSE 
\STATE $x_j\gets1$ 
\ENDIF 
\STATE Schedule $j$ using List Scheduling with respect to the assignment variable $x_j$. \label{line:LG-sched}
\end{algorithmic} 
\end{algorithm} 
 
LG has then been improved into the algorithm \emph{Modified Greedy} (MG) to allocate more tasks to GPUs.
In this algorithm, detailed in \Cref{alg:MG}, the tasks that have been assigned to GPUs by MG but not by LG are stored into a set $R$.
Any new task whose processing time on CPU is larger than a lower bound on the makespan needed to schedule the $R$ tasks and this new task on GPU is then assigned to GPU.
This modification ensures that the competitive ratio is not impacted by large tasks.
Indeed, Imreh proves that MG is $(4-2/m$)\,-\,competitive~\cite{Imreh2003}.
The principle of this algorithm is similar to Al4, presented in \Cref{sssec:al4}.
Because of the use of the set $R$, MG has a slightly better
competitive ratio ($4-2/m$ versus $4$), but its proof requires
considering more cases and to perform a tighter analysis.
Hence, we only sketch in this survey the proof of the competitive ratio of Al4 and not the one of MG.
The time complexity of scheduling any task is the same as with LG
because the computation on Line~\ref{line:MG-comp} can be performed
incrementally in constant time.
 
\begin{algorithm} 
\caption{Modified Greedy (MG)~\cite{Imreh2003}}\label{alg:MG} 
\begin{algorithmic}[1] 
\STATE $R\gets\emptyset$
\STATE Upon arrival of task $j$: 
\IF {${\overline{p_j}}/{m} \geq {\underline{p_j}}/{k}$} 
\STATE $x_j\gets 0$ 
\ELSIF {$\overline{p_j} \geq \max\left(\max_{i\in R\cup \{j\}} \underline{p_i}~,~  \sum_{i\in R\cup \{j\}} \underline{p_i}/k\right)$} \label{line:MG-comp}
\STATE $x_j\gets 0$ 
\STATE Add task $j$ to $R$.
\ELSE 
\STATE $x_j\gets1$ 
\ENDIF 
\STATE Schedule $j$ using List Scheduling with respect to the assignment variable $x_j$.
\end{algorithmic} 
\end{algorithm}

\subsubsection{\emph{Al4} and \emph{Al5}}\label{sssec:al4}

Chen et al.~\cite{Guochuan} proposed several algorithms, both for the general case and for the two special cases where $k = m$ and $k = 1$.
For the general case, the first algorithm \emph{Al4} presented in \Cref{alg:al4} combines two decision rules to assign a task to a CPU or a GPU and then schedules the task using List Scheduling.
Notice that $\underline{\tau}$ on Line~\ref{line:al4rule1} of the algorithm denotes the earliest moment when at least one GPU is idle.
The time complexity of this algorithm is then $O(\log(m))$, including both
the computation of $\underline{\tau}$ and the
last scheduling phase.
The algorithm applies the following two rules:
\begin{description} 
\item[Rule 1] On Line~\ref{line:al4rule1}, a task is assigned to GPUs if its running time on CPUs is larger than its completion time on GPUs; 
\item[Rule 2] On line~\ref{line:al4rule2}, a task is assigned to CPUs if its \emph{average weight} is larger on GPUs, with a similar rule than the LG algorithm.
\end{description} 
 
\begin{algorithm} 
\caption{Al4~\cite{Guochuan}}\label{alg:al4} 
\begin{algorithmic}[1] 
\STATE Upon arrival of task $j$:
\IF { $\overline{p_j} \geq \underline{\tau} + \underline{p_j}$} \label{line:al4rule1}
  \STATE $x_j \gets 0$
\ELSIF {$\overline{p_j}/m \leq \underline{p_j}/k$}\label{line:al4rule2}
  \STATE $x_j \gets 1$
\ELSE
  \STATE $x_j \gets 0$
\ENDIF
\STATE Schedule $j$ using List Scheduling with respect to the assignment variable $x_j$.
\end{algorithmic} 
\end{algorithm} 
 
\begin{theorem} \label{thm:al4} 
Al4 (\Cref{alg:al4}) achieves a competitive ratio of 4.
\end{theorem} 
 
In the proof of \Cref{thm:al4}, the authors compare on which type of machine a task is placed in the output schedule of \emph{Al4} (denoted by $S_{Al4}$ with makespan $C_{max}$) and the optimal schedule (denoted by $S_{OPT}$ with makespan $C_{max}^*$).
More precisely, the set of tasks is partitioned into five disjoint subsets as follows:
\begin{itemize} 
  \item[-] $\Lambda_C$ (resp. $\Lambda_G$): Set of tasks scheduled on CPUs (resp. GPUs) for both $S_{Al4}$ and $S_{OPT}$
  \item[-] $U_G$: Set of tasks scheduled on GPUs in $S_{Al4}$ by Rule 1 but on CPUs in $S_{OPT}$
  \item[-] $V_C$ (resp. $V_G$): Set of tasks scheduled on CPUs (resp. GPUs) in $S_{Al4}$ by Rule 2 but on GPUs (resp. CPUs) in $S_{OPT}$
\end{itemize} 
 
Let $\lambda_C$, $\lambda_G$, $u_G$, $v_C$ and $v_G$ denote the total processing times, according to $S_{Al4}$, of tasks in sets $\Lambda_C$, $\Lambda_G$, $U_G$, $V_C$ and $V_G$, respectively. Moreover,  
\begin{equation} 
C_{max} \leq \max \left\{ \frac{\lambda_C + v_C}{m} + \overline{p_{max}}, 
\frac{\lambda_G + u_G + v_G}{k} + \underline{p_{max}} 
\right\} \label{eqn:al4cmax1} 
\end{equation} 
where $\overline{p_{max}}$ (resp. $\underline{p_{max}}$) is the processing time of the largest task on CPUs (resp. GPUs) in $S_{Al4}$.
 
From the relations between processing times in the decision rules, it is possible to derive the following bounds (more details can be found in the original paper~\cite{Guochuan}): 

\begin{align*} 
\frac{u_G}{k} &\leq C_{max}^* \\ \frac{\lambda_C}{m} + \frac{v_G}{k} 
&\leq C_{max}^* \\ \frac{\lambda_G}{k} + \frac{v_C}{m} &\leq C_{max}^* 
\end{align*} 
\medskip 
 
It remains now to bound $\overline{p_{max}}$ and $\underline{p_{max}}$.
Let $j$ denote the task with processing time $\overline{p_{max}}$.
If $j$ is also on a CPU in the optimal solution, then $\overline{p_{max}} \leq C_{max}^*$.
If $j$ is on a GPU in $S_{OPT}$, then $\underline{p_j} \leq C_{max}^*$ and since it has not been scheduled on CPUs by Rule 1, then
 
\[ \overline{p_{max}} = \overline{p_j} \leq \frac{\lambda_G + u_G + v_G}{k} + \underline{p_j} \leq \frac{\lambda_G + u_G + v_G}{k} + C_{max}^* \] 
\medskip 
 
Let $j'$ denote the task with processing time $\underline{p_{max}}$.
If $j'$ is also on a GPU in $S_{OPT}$, $\underline{p_{max}} \leq C_{max}^*$.
If $j'$ is on a CPU in $S_{OPT}$, then we can derive from the bounds
\[\underline{p_{max}} = \underline{p_{j'}} \leq \overline{p_{j'}} \leq C_{max}^* \] 
\medskip 
 
Finally, by plugging these bounds into \Cref{eqn:al4cmax1}, we obtain $C_{max} \leq 4 C_{max}^*$, which concludes the proof.
 
The authors also propose a refinement of this algorithm by adding a third decision rule and by adding coefficients to the rules so as to achieve a better load balancing between the two sets of resources.
This improved algorithm achieves a competitive ratio at most $3.85$ with the same time complexity.
We consider this improved algorithm, denoted by \emph{Al5}, instead of \emph{Al4} in the experimental session since it provides a slightly better competitive ratio.
 
Moreover, for the special case where $k = m$, they propose a 3-competitive algorithm that can even be improved to be $(1+\sqrt{3})$-competitive by adding another decision rule and coefficients in a similar way as for the general case.
For the one-sided case, where $k=1$, a 3-competitive algorithm is also provided.
These special case algorithms have a time complexity similar to the one of Load Greedy (\Cref{sssec:LGMG}).

\section{Approximation Algorithms and Heuristics for Tasks with Precedence Constraints}\label{sec:prec}

Many real-life parallel computing applications consist of tasks linked with precedence relations induced by data dependencies, which complexifies the search for efficient schedules.
In this section, we review scheduling algorithms that have been proposed for applications with precedence constraints on heterogeneous platforms.

\subsection{Off-line Setting for Tasks with Precedence Constraints}\label{ssec:precOFF}

We first present existing strategies for the off-line case, that is, when both the structure of the precedence constraints and the cost of the tasks are completely known beforehand.

\subsubsection{HEFT}\label{sssec:HEFT}

When scheduling tasks with precedence constraints on heterogeneous resources, \emph{Heterogeneous Earliest Finish Time} (HEFT)~\cite{HEFT2002} is an unavoidable heuristic.
Despite its apparent simplicity and a non-constant approximation ratio~\cite{DP43,europarPaper}, it performs well in most cases and is thus used as a reference heuristic in many recent scheduling studies.

HEFT, presented in \Cref{alg:HEFT}, extends the Earliest Finish Time principle to heterogeneous platforms.
It consists in two steps: (i) tasks ranking and (ii) resources selection.
To rank the tasks, HEFT generalizes the concept of bottom-level to heterogeneous platforms by using the average computation time of a task on all machines (and the average communication time among two machines).
The rank of task $j$ is defined as
$$ \mathit{rank}(j) = w_j + \max_{i\in \Gamma^-(j)} (c_{j,i} +
\mathit{rank}(i)),
$$ 
where $w_j$ is the average processing cost of task $j$ ($(m\overline{p_j} + k \underline{p_i})/(m+k)$ in our case) and $c_{j,i}$ the average communication cost of edge $(j,i)$ (assumed to be negligible in our setting).
Tasks are then sorted by decreasing values of $\mathit{rank}(j)$, which provides a topological ordering.
Then, ready tasks are considered in this order to be scheduled on the resources.
The first (ready) task according to this ordering is scheduled on the resource that is able to complete this task the soonest, following EFT principle.
During this second phase, precedence constraints are considered (a task cannot start until all its predecessors are completed) as well as resource availability: HEFT uses an insertion-based strategy, i.e., all possible idle slots on all possible resources are considered to schedule a task.
Its time complexity is in $O(n^2)$ and results from the time to
compute the ranks and this last insertion-based strategy.

While HEFT is designed as an off-line algorithm, its task sorting policy has been transposed to dynamic schedulers that only consider tasks at runtime once all their predecessors have been processed.
This is typically the case of the DMDA scheduler of StarPU runtime~\cite{AugThiNamWac11CCPE}.

\begin{algorithm}
\caption{Heterogeneous Earliest Finish Time
  (HEFT)~\cite{HEFT}}\label{alg:HEFT}
\begin{algorithmic}[1]
\STATE Compute the rank of each task $j \in \mathcal{T}$.
\STATE $L \gets \mathcal{T}$ sorted by decreasing rank of the tasks.
\FOR{each task $j \in L$}
\STATE Schedule task $j$ on the machine where it will be finished the
earliest with an insertion-based strategy.
\ENDFOR
\end{algorithmic}
\end{algorithm}

\subsubsection{HLP}
\label{sssec:hlp}

\emph{Heterogeneous Linear Program} (HLP), presented in \Cref{alg:HLP} is the first algorithm with a proved constant approximation ratio for the problem of scheduling tasks with precedence constraints on two types of resources~\cite{HLP}.
It relies on the solution of the linear program (in rational numbers) $LP_{prec}$ (defined in \Cref{sssec:precLB}) combined with a rounding strategy of its fractional solution, that decides on the allocation of each task.
This step is followed by a variant of List Scheduling adapted to the case of two resource types, where tasks have been sorted according to the \emph{Earliest Starting Time} (EST) strategy.
In addition to solving the linear program, the time complexity of the
remaining operations is $O(n\log(n)+n\log(m))$ where the first term is
for sorting tasks by their earliest starting time and the second is
for finding the first available resource.

After the resolution of the relaxed version of $LP_{prec}$, $x_j$ variables end up with fractional values.
The rounding rule applied to the fractional solution is common: it consists in setting $x_j$ to $1$ if the fractional value is $\geq 1/2$ and to $0$ otherwise.
This leads to a 2-approximation of the fractional allocation with respect to the optimal one, as shown in Lemma 2 of the original paper~\cite{HLP}.

\begin{algorithm}
\caption{Heterogeneous Linear Program (HLP)~\cite{HLP}}\label{alg:HLP}
\begin{algorithmic}[1]
\STATE Solve the linear program $LP_{prec}$ over the rational numbers.
\STATE Let $\tilde{x}_j$ be the (fractional) value of the assignment of the variable $j$ in an optimal solution of $LP_{prec}$.
\FOR {each task $j$}
  \IF {$\tilde{x}_j \geq 1/2$ }
    \STATE $x_j \gets 1$
  \ELSE
    \STATE $x_j \gets 0$
  \ENDIF
\ENDFOR
\STATE $S_{A} \gets \emptyset$
\WHILE {${S_{A} \neq \mathcal{T}}$}
  \STATE Select the ready task $j$ that has the earliest starting time (EST policy).
  \STATE Schedule $j$ using List Scheduling with respect to the assignment variable $x_j$.
  \STATE Add $j$ to $S_{A}$.
\ENDWHILE
\end{algorithmic}
\end{algorithm}

\begin{theorem}\label{thm:HLP}
HLP (\Cref{alg:HLP}) is a 6-approximation algorithm, and this ratio is tight.
\end{theorem}

Let us sketch the proof of the first part of Theorem~\ref{thm:HLP} (more details may be found in the original paper~\cite{HLP}).
The analysis follows the principle of List Scheduling recalled in \Cref{ssec:preliminaries}.
The main difference here is that the whole span $I=[0,C_{max})$ of time slots is decomposed into 3 subsets of intervals instead of $2$, namely $I_{CP}$, $\overline I$ and $\underline I$.
$I_{CP}$ includes the time slots when at least one CPU and one GPU are idle, while $\overline I$ (resp. $\underline I$) includes the time slots when all the CPUs (resp. GPUs) are fully occupied.
Since the intersection between $\overline I$ and $\underline I$ may be non-empty, the makespan is bounded above by the sum of the duration of the time slots in each of the three subsets.
Clearly, the overall length of $I_{CP}$ is upper bounded by the length of the critical path, while the length of $\overline I$ (resp. $\underline I$) is upper bounded by the average workload on the CPUs (resp. GPUs).
Moreover, rounding $x_j$ variables can at most double the objective value $C_{LP}$ of $LP_{prec}$. Thus, as $C_{LP}$ is a lower bound of the optimal feasible makespan we get
\[ C_{max} \leq |I_{CP}| + |\overline I| + |\underline I| \leq 6C_{LP} \leq 6C_{max}^* . \]

\bigskip
This algorithm has been further studied by Amaris et al.~\cite{europarPaper}.
They propose a variant of the scheduling policy called \emph{Ordered List Scheduling} (OLS).
In this policy, a ranking similar to HEFT~\cite{HEFT2002} is computed for each task and the list of tasks is sorted in decreasing order of the ranks before using List Scheduling.
The resulting algorithm is called HLP-OLS in reference to HLP-EST, the original algorithm based on EST policy.

The authors show that, although the approximation ratio of HLP-OLS is also 6, OLS policy performs better in practice.
Moreover, they propose a worst-case example for HLP achieving a ratio of $6 - O(\frac{1}{m})$ whatever the scheduling policy applied during the second phase, which proves the tightness of the approximation ratio.

\subsection{On-line Setting for Tasks with Precedence Constraints}\label{ssec:precON}

Finally, let us review proposed scheduling strategies for the most difficult problem, where tasks with precedence constraints are discovered by the scheduler as they are made available by the completion of the tasks they depend on.
Note that the heuristic PG (also named ECT) described in Section~\ref{sssec:LGMG} may well be applied to this scenario.
As for the independent tasks case, the competitive ratio of this algorithm is at least $\lfloor m/k\rfloor$.

\subsubsection{ER-LS}
\label{sssec:erls}

\emph{ER-LS}~\cite{concurrency} differs from $\emph{Al4}$ (presented
in~\Cref{sssec:al4}) in its second rule that diminishes the role of
the number of each resources by considering their square roots.
This new set of rules gives more importance to the acceleration factor
and is presented in \Cref{alg:erls}.

\begin{algorithm}
\caption{ER-LS~\cite{concurrency}}\label{alg:erls}
\begin{algorithmic}[1]
\STATE Upon arrival of task $j$:
\IF { $\overline{p_j} \geq \underline{\tau} + \underline{p_j}$}
  \STATE $x_j \gets 0$
\ELSIF {$\overline{p_j}/\sqrt{m} \leq \underline{p_j}/\sqrt{k}$}
  \STATE $x_j \gets 1$
\ELSE
  \STATE $x_j \gets 0$
\ENDIF
\STATE Schedule $j$ using List Scheduling with respect to the assignment variable $x_j$.
\end{algorithmic}
\end{algorithm}

\begin{theorem} \label{thm:erls}
ER-LS (\Cref{alg:erls}) achieves a competitive ratio of $4\sqrt{\frac{m}{k}}$.
\end{theorem}

The proof of \Cref{thm:erls} is similar to the one of \Cref{thm:al4}, where $p_{max}$ is replaced by the length of the critical path in the schedule, and is therefore omitted.
The full proof can be found in the original paper~\cite{concurrency}.
The result is established by proving separately the three following inequalities on the schedule produced by ER-LS:
$$ C_{max} \leq \frac{\overline{W}}m + \frac{\underline{W}}{k} + CP \quad;\quad CP \leq \sqmkl\ C_{max}^* \quad;\quad \frac{\overline{W}}m + \frac{\underline{W}}{k} \leq 3\, \sqmkl\ C_{max}^*.$$

\subsubsection{QA and mixed-ECT-QA}
\label{sssec:qa}
\newcommand{\qa}{\emph{QA}\xspace}
\newcommand{\algomix}{\emph{Mixed-ECT-QA}\xspace}

\emph{Quick Allocation} (QA)~\cite{europar18} can be seen as a simplification of $\emph{ER-LS}$ algorithm presented in \Cref{sssec:erls}, that exhibits a better competitive ratio.
The first decision rule for $\emph{ER-LS}$ is removed and the new set
of rules is presented in \Cref{alg:qa} leading to the time complexity
of scheduling a task $O(\log(m))$ as with PG and LG\@.
The intuitive goal of this first rule was to complete the first large tasks as early as possible, possibly at the price of the global optimality.
This behavior was necessary in \emph{Al4} algorithm  (on which ER-LS is based) as the target approximation ratio was smaller than $\sqrt{m/k}$.
We show here that this rule is superfluous when scheduling tasks with precedence relations and leads to a larger competitive ratio.

\begin{algorithm}
\caption{Quick Allocation (\qa)~\cite{europar18}}\label{alg:qa}
\begin{algorithmic}[1]
  \STATE Upon arrival of task $j$:
  \IF {$\overline{p_j}/\sqrt{m} \leq \underline{p_j}/\sqrt{k}$}
    \STATE $x_j \gets 1$
  \ELSE
    \STATE $x_j \gets 0$
  \ENDIF
  \STATE Schedule $j$ using List Scheduling with respect to the
assignment variable $x_j$.
\end{algorithmic}
\end{algorithm}

\begin{theorem} \label{thm:qa}
\qa (\Cref{alg:qa}) achieves a competitive ratio of
$2\sqrt{\frac{m}{k}}+1-\frac{1}{\sqrt{mk}}$.
\end{theorem}

The proof of \Cref{thm:qa}, detailed in the original publication~\cite{europar18}, is adapted from the one of \Cref{thm:erls} with tighter inequalities and is therefore omitted here.
The following inequalities are established in the proof:

$$ C_{max} \leq \frac{\overline{W}}m + \frac{\underline{W}}{k} + \left(1-\frac1m\right) CP \quad;
  \quad CP \leq \sqmkl\ C_{max}^* \quad;
  \quad \frac{\overline{W}}m + \frac{\underline{W}}{k} \leq \left(\sqmkl+1\right) C_{max}^*.
$$

\medskip
The competitive ratio of the \qa algorithm is almost tight as stated in the following theorem, proved in the original paper~\cite{europar18}.
\begin{theorem} \label{th:QALSESTlb}
The competitive ratio of the \qa algorithm is at least $\left(2\sqmk+1-\frac{1}k\right)$.
\end{theorem}

The main idea to build the lower bound example is to combine many short independent tasks of acceleration factor $\sqrt{m/k}+\varepsilon$ with a single long task of acceleration factor $\sqrt{m/k}-\varepsilon$, which depends on a small task that needs to be run on GPU.
The algorithm QA schedules the first set of tasks on the GPU and the single task on CPU afterwards, losing a time factor of $\sqrt{m/k}$ in both cases compared to the reversed allocation.

\medskip

An advantage of \emph{ER-LS} over \qa is that it better schedules some \emph{easy} instances, such as single task instances, thanks to its first rule.
The idea of relying on \qa in order to obtain a good competitive ratio while improving the scheduling decisions for easy instances is the motivation of the algorithm \algomix,  introduced by the same authors~\cite{europar18} and presented in \Cref{alg:mix}.
This algorithm is parameterized by a factor $\gamma$ and works as follows.
Initially, it takes the same decisions as the ECT algorithm, presented in \Cref{sssec:LGMG}.
If the achieved makespan is at least $\gamma$ times longer than the one that would be achieved by \qa, then all the subsequent scheduling decisions are those \qa would have made.
Intuitively, this algorithm initially works like ECT since ECT performs well on many easy instances, but if the instance is identified as being difficult, then it switches to \qa algorithm which has a better competitive ratio.
\algomix is therefore as efficient as ECT on easy instances, but it
achieves a smaller competitive ratio in the worst case. The time
complexity to schedule a task is dominated by the computation of the
schedule that would be achieved by \qa, which is $O(n\log m)$.

\begin{theorem}
\algomix (\Cref{alg:mix}) achieves a competitive ratio of
$(\gamma+1)(2\sqrt{\frac{m}{k}}+1)$.
\end{theorem}

\begin{algorithm}
\caption{\algomix$(\gamma)$~\cite{europar18}\label{alg:mix}}
\begin{algorithmic}[1]
\STATE $\mathit{StayECT} \gets \mathit{true}$
\STATE Upon arrival of task $j$:
\IF{$\mathit{StayECT}$}
  \STATE $C_{EFT}\gets $ makespan obtained by scheduling task $j$ as EFT
  \STATE $C_{QA} \gets$ Makespan that QA would have obtained on the whole known graph
  \STATE $\mathit{StayECT} \gets (C_{ECT} \leq \gamma C_{\qa})$
\ENDIF
\IF{$\mathit{StayECT}$}
  \STATE Schedule $T_i$ (as soon as possible) on the resource which is able to complete it the earliest.
\ELSE
  \STATE Schedule $T_i$ (as soon as possible) on CPU if $\overline{p_i} / \underline{p_i} \leq \sqmkl$ and on GPU otherwise.
\ENDIF
\end{algorithmic}
\end{algorithm}

\section{Experiments}\label{sec:experiments}

This section presents experimental results to compare the behavior of all algorithms discussed in this paper in practice.
All algorithms used in this section have been implemented in C++ as part of the \texttt{pmtool} project~\cite{pmtool}, and linear programs are solved using IBM CPLEX v12.7. All input data and experimental analysis are available in the companion repository: \url{https://hal.inria.fr/hal-02159005}.

\subsection{Independent Tasks}
\label{ssec:expe.independent}

\subsubsection{Algorithms}


Almost all the algorithms presented in Section~\ref{sec:independent} have been implemented.
In particular, this includes the following off-line strategies:
\begin{itemize}
\item The Sorted ECT algorithm, which considers tasks with highest \emph{average} execution time first.
      This algorithm is actually equivalent to HEFT~\cite{HEFT} in the case of independent tasks.
\item HeteroPrio, as described in Section~\ref{sssec:heteroprio}.
\item BalancedEstimate and BalancedMakespan~\cite{BalancedEuropar}, as described in Section~\ref{sssec:balanced} (their names have shortened to BalEst and BalMks in the plots).
\item CLB2C~\cite{CLB2C}, as described in Section~\ref{sssec:clb2c}.
\item The algorithms based on dual approximation technique:
\begin{itemize}
  \item DualHP~\cite{fastbio}, as described in Section~\ref{sssec:dualhp}.
  \item DP$_{2}$~\cite{DP43} and DP$_{\frac{3}{2}}$, as described in Section~\ref{sssec:DP}.
        In practice, to accelerate the execution times, these algorithms have not been implemented using dynamic programming, but rather using  the corresponding integer programming formulation and CPLEX.
  \end{itemize}
\end{itemize}

Algorithms based on (relaxations of the) linear programming were also implemented:
\begin{itemize}
\item The LP algorithm solves the Integer Linear Programming formulation with CPLEX, with a gap of at most 10\% (the solver stops when it finds an integer solution provably within 10\% of the optimal solution).
      This algorithm has no guarantee of polynomial execution time.
\item Round denotes the algorithm that solves the rational relaxation of the Linear Programming formulation described in~\ref{sssec:indepOffLB}, and then rounds the solution as described by Tarplee et al.~\cite{Tarplee201576}.
      Round algorithm is designed for an arbitrary number of resource types; In the case of two resources, this corresponds to rounding to the closest integer value.
\end{itemize}

Moreover, the MinMin algorithm~\cite{MinMin}, designed for scheduling independent tasks on heterogeneous platforms, was also added in our set of tested algorithms.
MinMin selects, over all unscheduled tasks, the one with the minimum expected completion time over all machines and schedules it on the corresponding machine.
This process is repeated while unscheduled tasks remain.

We also included the on-line algorithms ECT, LG and MG depicted in Section~\ref{sssec:LGMG}, and Al5 from Section~\ref{sssec:al4}, with its default parameters set as in the associated publication~\cite{Guochuan} (Al4 was not implemented as it has a slightly lower competitive ratio).

Furthermore, since the exact computation of an optimal solution is costly, we used for each instance
the lower bounds given in Section~\ref{sssec:indepOffLB} to normalize the makespan of all above algorithms.
In particular, we computed the area bound, which is the optimal value of the rational solution of the Linear Program solved by Round.

\subsubsection{Benchmarks and Results}

We consider two different families of instances.
The first family consists of randomly generated instances, which allows us to explore a wide variety of scenarios and assess an average behavior of all algorithms.
The second family contains benchmarks from real-life linear algebra kernels, in order to analyze the practical behavior of algorithms.

\paragraph{Random instances}
These tasks are generated with the same procedure as in the original publication of BalancedEstimate and BalancedMakespan~\cite{BalancedEuropar}: each task duration on a CPU follows a Gamma distribution of expected value 15, and durations on GPU follow a Gamma distribution of expected value 1.
The durations of the different tasks are independent, and the respective durations on both types of resources for a given task are independent as well.
The coefficient of variation of these distributions can be either 0.2 (low) or 1 (high).
This yields to four different cases.
Each instance contains 300 tasks, and for each setting 100 instances are generated.
The number of CPUs is either 10 or 40, and the number of GPUs is either 2 or 8.

\begin{figure}
  \begin{center}
    \includegraphics[width=\linewidth]{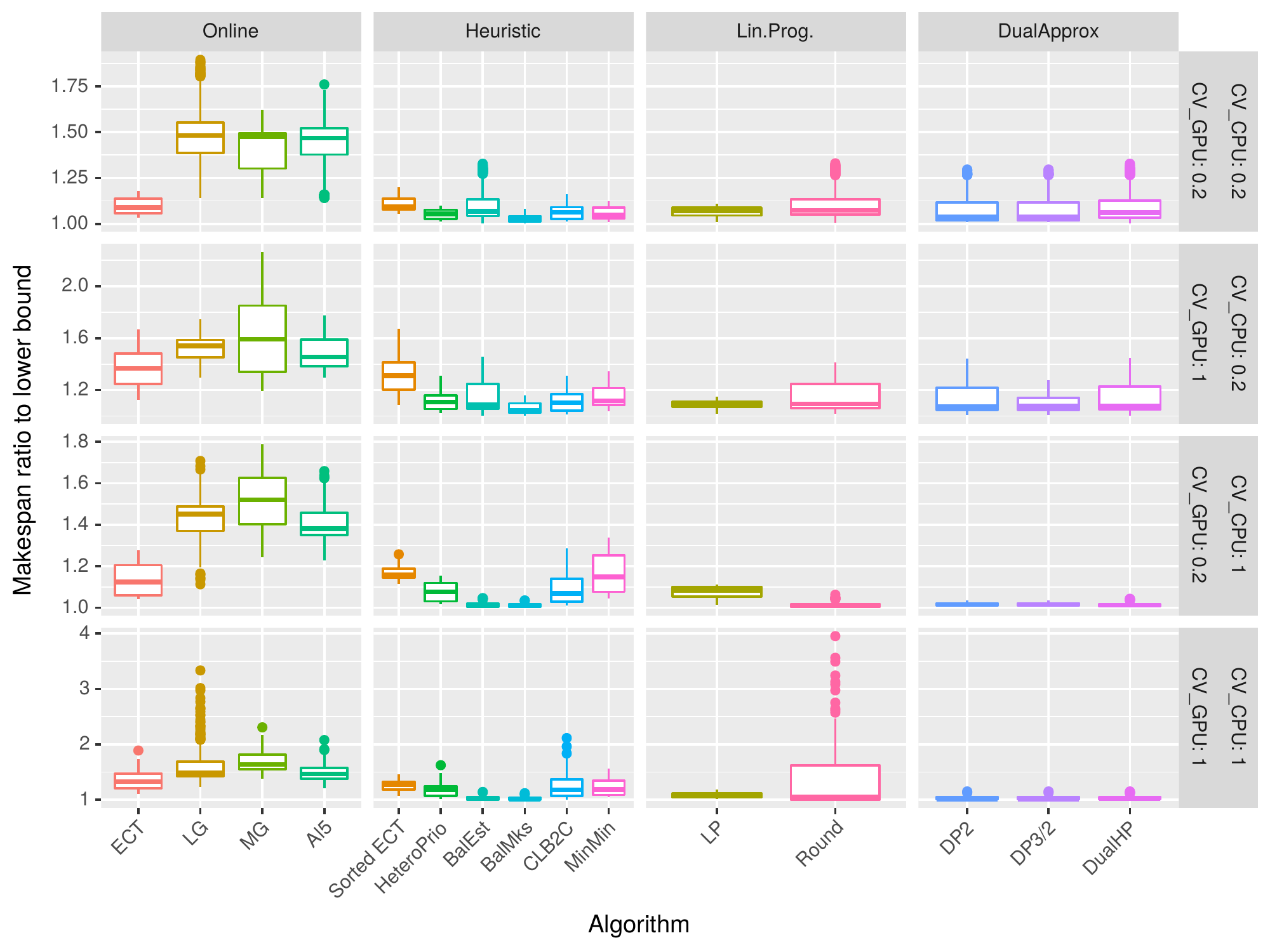}
  \end{center}
  \caption{Experimental results for the Random Independent case.
    Column labels show the type of algorithms, row labels show the coefficients of variation of the distributions.}
  \label{fig:expe:randomindep}
\end{figure}

\Cref{fig:expe:randomindep} depicts the quality of the schedules produced by all algorithms.
As mentionned above, the quality of a schedule is assessed through the ratio of the makespan to a lower bound of the optimal makespan.

On this plot (and on all the following plots in this section), the results are represented using boxplots: for a group of values, the bottom and the top of the rectangle correspond to the first and third quartile, and the line inside corresponds to the median measurement.
Whiskers extend to the extreme values, but no further than $1.5$ times the inter-quartile range (which is the height of the rectangle).
Values beyond the end of the whiskers are plotted individually.
Since the results do not exhibit a significant correlation with the number of resources, all the results for a given case are grouped together.
We can make the following observations:

\begin{itemize}
\item As expected, on-line algorithms achieve worse performance than off-line algorithms.
ECT is on par with the worst off-line algorithms; Al5 achieves the best results among all on-line algorithms with a performance guarantee.
\item Sorted ECT and MinMin are significantly worse than other off-line algorithms.
In particular, sorting the tasks does not significantly improve the performance of ECT.
On another hand, BalancedMakespan consistently achieves the best performance in all cases.
HeteroPrio and CLB2C are based on very close ideas, and indeed behave very similarly.
\item All algorithms based on dual approximation exhibit similar behaviors, despite their different performance guarantees.
On instances with low variation on the distribution of execution times on the CPUs, their performance is significantly worse than with higher variation.
BalancedEstimate  exhibits the same behavior.
In this case, we believe that this comes from a low quality of the approximation of the makespan of an allocation as the ratio of the total load to the number of resources.
\item There is a high variation of the execution times of tasks on the GPUs (CV\_GPU=1) for the most difficult instances: the ratios for all algorithms are higher than in the low variation cases.
\end{itemize}

\begin{figure}
  \begin{center}
    \includegraphics[width=\linewidth]{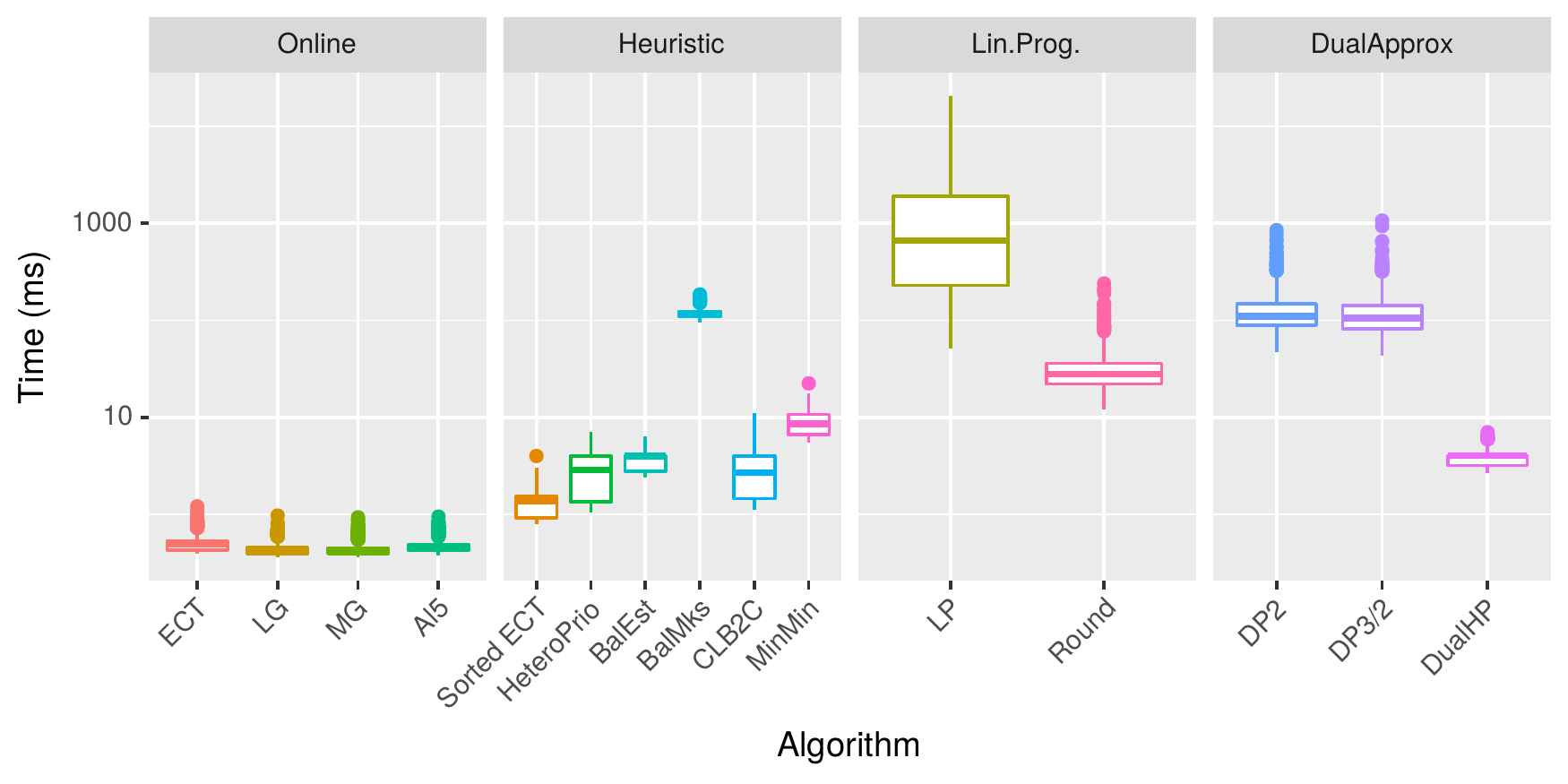}
  \end{center}
  \caption{Computation time for the Random Independent case.
    Column labels show the number of GPUs, row labels show the number of CPUs.}
  \label{fig:expe:randomindep:time}
\end{figure}

\Cref{fig:expe:randomindep:time} presents the running times of all algorithms.
Since the running time does not depend significantly on the instance types, all the results are grouped together, and a logscale is used in order to properly display the results.
As expected, on-line algorithms are the fastest, and the longest running times correspond to algorithms relying on the solution of an LP.
The good performance of BalancedMakespan actually requires a larger computation time than for other heuristics, since it requires recomputing a whole schedule at each allocation decision.
The dual approximation approach is also costly, because it requires solving the allocation problem (either by Dynamic or Linear Programming).

\paragraph{Linear Algebra Kernels}

We consider instances introduced by Beaumont et al.~\cite{HeteroPrioJournal} in a previous work on HeteroPrio.
In order to obtain a representative mix of different kernels, applications from the chameleon suite~\cite{chameleon} (Cholesky, LU, QR) have been executed on the \texttt{sirocco} platform with tile size 960.
Each of these applications consists in many calls to a few linear algebra kernels, which correspond to the individual tasks of the applications.
Although there are precedence constraints among these tasks in actual applications, we first remove them in order to test the quality of all discussed scheduling strategies for independent tasks.
We measure the average running time of each kernel, as well as the number of times it is ran in each application, for a number of ($960 \times 960$) tiles varying from 6 to 20.
This results in a total number of tasks in these instances varying from 124 to 2874.
As before, these instances are simulated on platforms with either 10 or 40 CPUs, and with either 2 or 8 GPUs.

\begin{figure}
  \begin{center}
    \includegraphics[width=\linewidth]{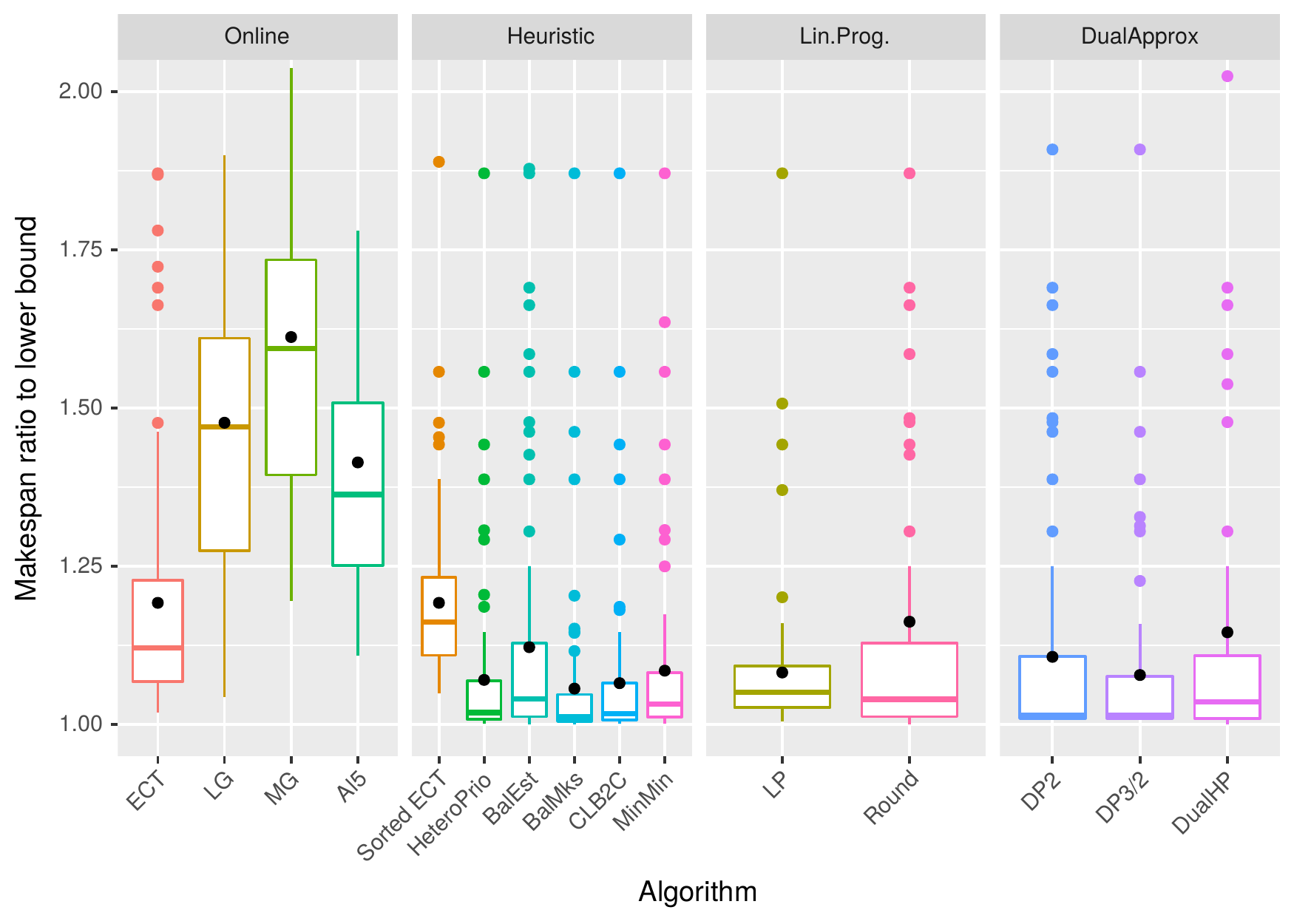}
  \end{center}
  \caption{Experimental results for the Linear Algebra Independent
    case.}
  \label{fig:expe:bordeauxindep}
\end{figure}

\Cref{fig:expe:bordeauxindep} depicts the quality of the schedules.
Since the results are similar when considering different number of tiles and of resource, as well as different applications, we group all the results corresponding to the same given strategy in a single boxplot.
For better readability, plots are truncated beyond 2 (the only impacted algorithms are LG and Al5, for which the maximum value is 2.3, and MG, Round, and DualHP, for which the maximum value is around 3), and a black dot is added to show the average ratio.
Note that even though DualHP is a 2-approximation, its makespan is larger than twice the lower bound on a few instances, in particular when there are only a few tasks.
This is due to the fact that we compare with a lower bound and not with the real optimal solution as in the analysis of the approximation ratio.

We observe the same trends as with the previous benchmarks: on-line algorithms achieve poor performance except for ECT; CLB2C and HeteroPrio achieve very close results; BalancedMakespan produces the best schedules among all algorithms.
These instances allow us to further differentiate between the various dual approximation algorithms: the sophisticated formulations obtain better solutions in some cases.
Sorted ECT is, as expected, not well adapted to independent tasks.

\begin{figure}
  \begin{center}
    \includegraphics[width=\linewidth]{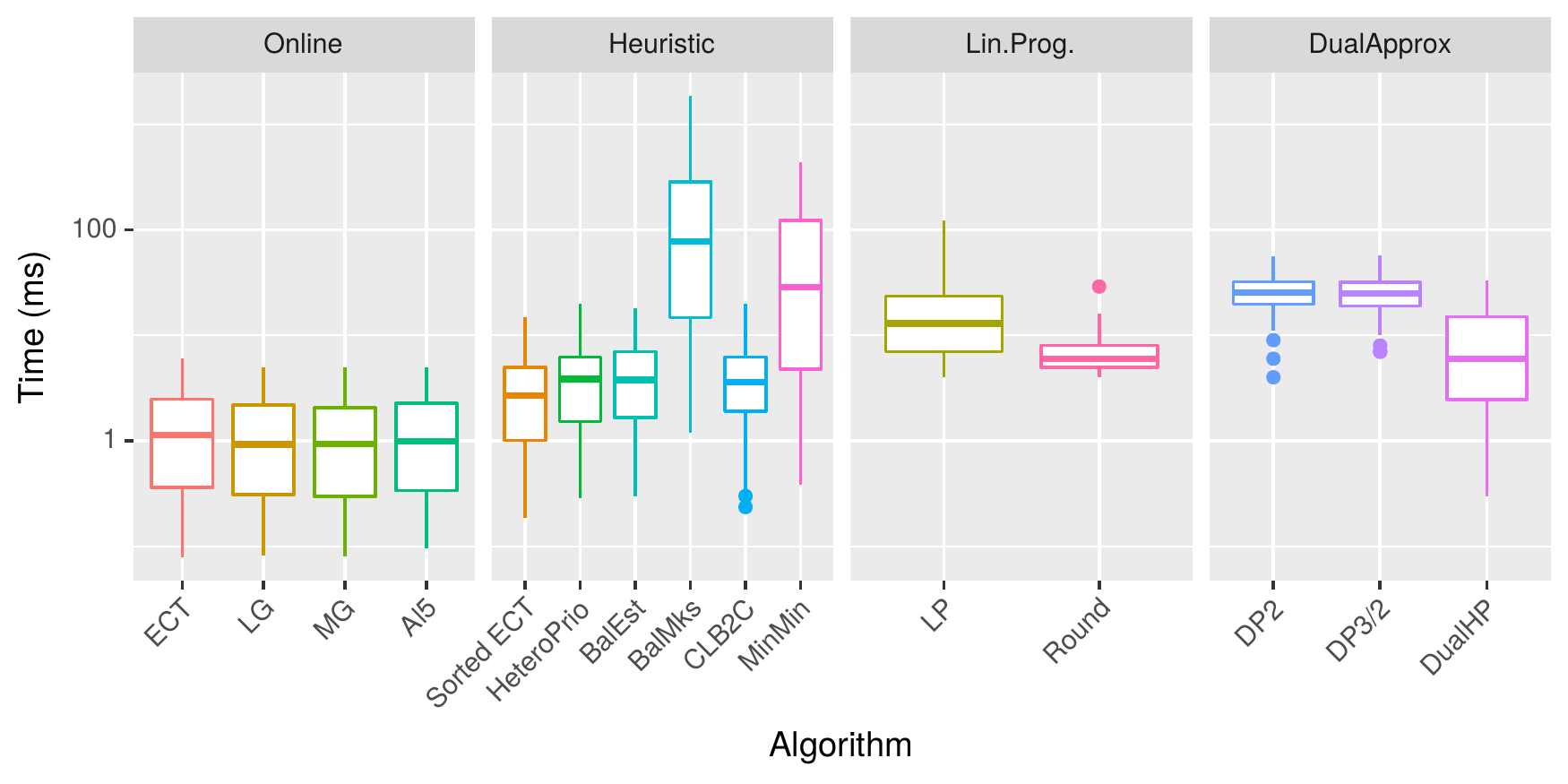}
  \end{center}
  \caption{Computation time for the Linear Algebra Independent case.}
  \label{fig:expe:bordeauxindep:time}
\end{figure}

\Cref{fig:expe:bordeauxindep:time} presents the running times of all algorithms, once again with a logarithmic scale.
Despite these instances have a larger number of tasks, the fact that they contain a small number (4 or 5 depending on the application) of task types induces shorter computation times than in the Random case depicted above.
Regardless, roughly the same behavior can be observed, except that the LP algorithm takes comparatively less time on this type of instances.

\subsection{Tasks With Precedence Constraints}
\label{ssec:expe.dag}

\subsubsection{Algorithms}

As in the case of independent tasks, almost all strategies presented in Section~\ref{sec:prec} have been implemented.
As for off-line algorithms, we consider:
\begin{itemize}
\item HEFT, as described in Section~\ref{sssec:HEFT}.
\item An off-line ECT variant (Off-line ECT) adapted to precedence constraints that computes priorities from the task graph instead of considering tasks in an arbitrary order.
The priority of a task corresponds to its distance (in terms of number of tasks) to any of its descendants.
\item HeteroPrio, as described in Section~\ref{sssec:heteroprio}, can easily be extended to the case with precedence constraints~\cite{HeteroPrioJournal}: whenever a resource is idle, a ready task is assigned to it from the list of ready tasks using the HeteroPrio policy.
If no task is ready, an idle GPU is allowed to spoliate a task from one of the CPUs if it can finish it earlier.
This version of HeteroPrio makes use of priorities (computed in a similar way as for HEFT) for two different purposes: (i) in order to break ties for tasks with the same acceleration factor, and (ii) in order to decide which task a GPU spoliates when it is idle.
\item The HLP algorithm, as described in Section~\ref{sssec:hlp}, has been implemented in two flavors: as described in Algorithm~\ref{alg:HLP} (a $6$-approximation) and with an additional spoliation strategy.
Indeed, it makes sense to include spoliation in the list scheduling phase of this algorithm: if some GPU is idle while there exists a task assigned to the CPU in the assignment phase, then the GPU is allowed to spoliate this task if it can finish it earlier.
\end{itemize}

We also implemented the on-line algorithms ECT (as described in \Cref{sssec:LGMG}), ER-LS (as described in \Cref{sssec:erls}), which is an extension of Al4, and QA (as described in \Cref{sssec:qa}), which is a simplification of ER-LS with stronger approximation ratio.

For each instance, we use as a lower bound the rational solution of the Linear Program presented in Section~\ref{sssec:precLB}.
This lower bound is used to normalize the makespan of all algorithms, as we did in the independent tasks case.

\subsubsection{Benchmarks and Results}

In order to consider realistic instances, we use the applications from the chameleon suite~\cite{chameleon}.
However, in this section, we consider the applications described with their dependency graph.
The number of tiles varies from 4 to 60.
We consider that the number of CPUs is either 10 or 40, and that the number of GPUs is either 2 or 8.

\begin{figure}
  \begin{center}
    \includegraphics[width=\textwidth]{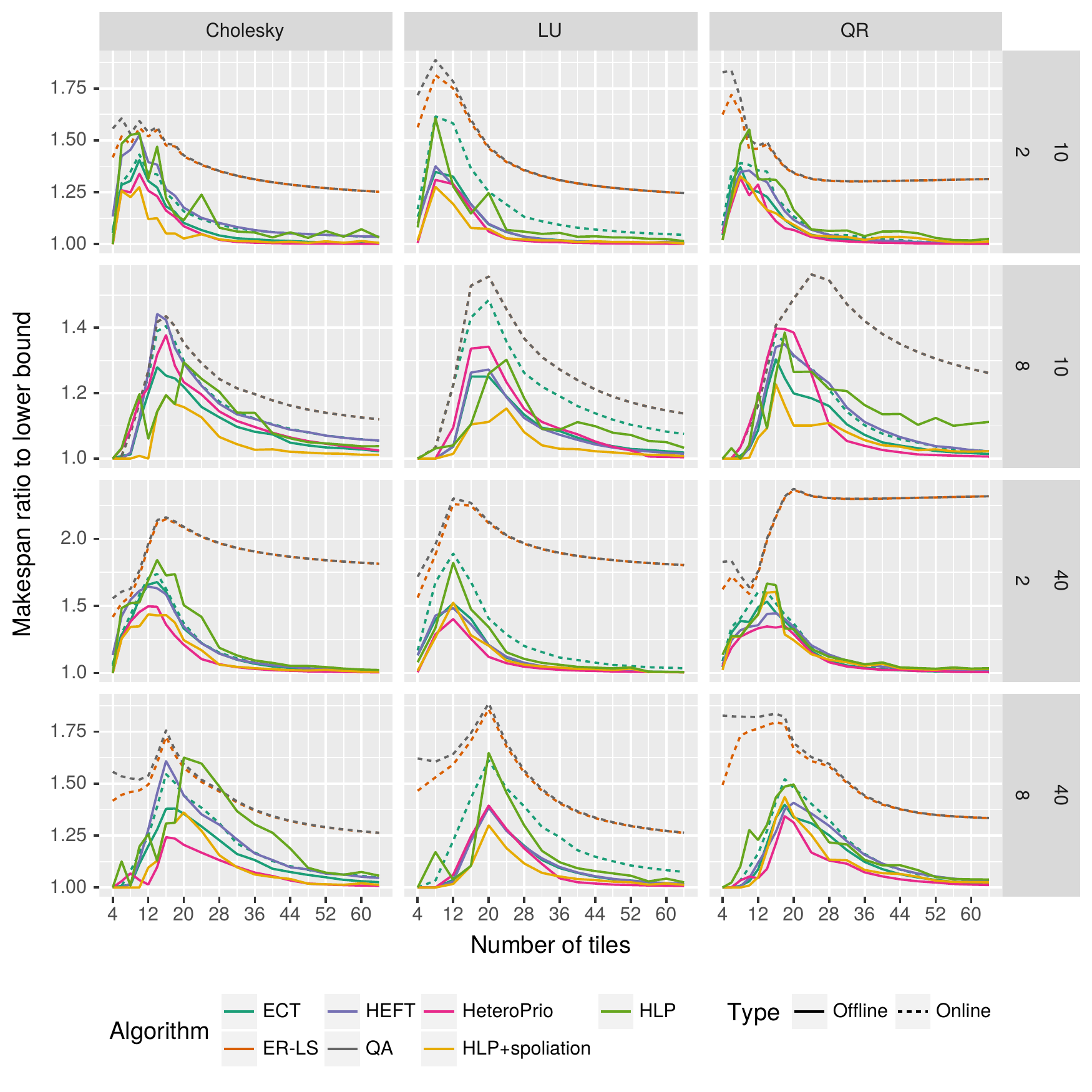}
  \end{center}
  \caption{Experimental results for Linear Algebra with
    dependencies. Column labels show the application, row labels show the
    number of CPUs (10-40) and GPUs (2-8).}
  \label{fig:expe:bordeaux}
\end{figure}

\begin{figure}
  \begin{center}
    \includegraphics[width=\textwidth]{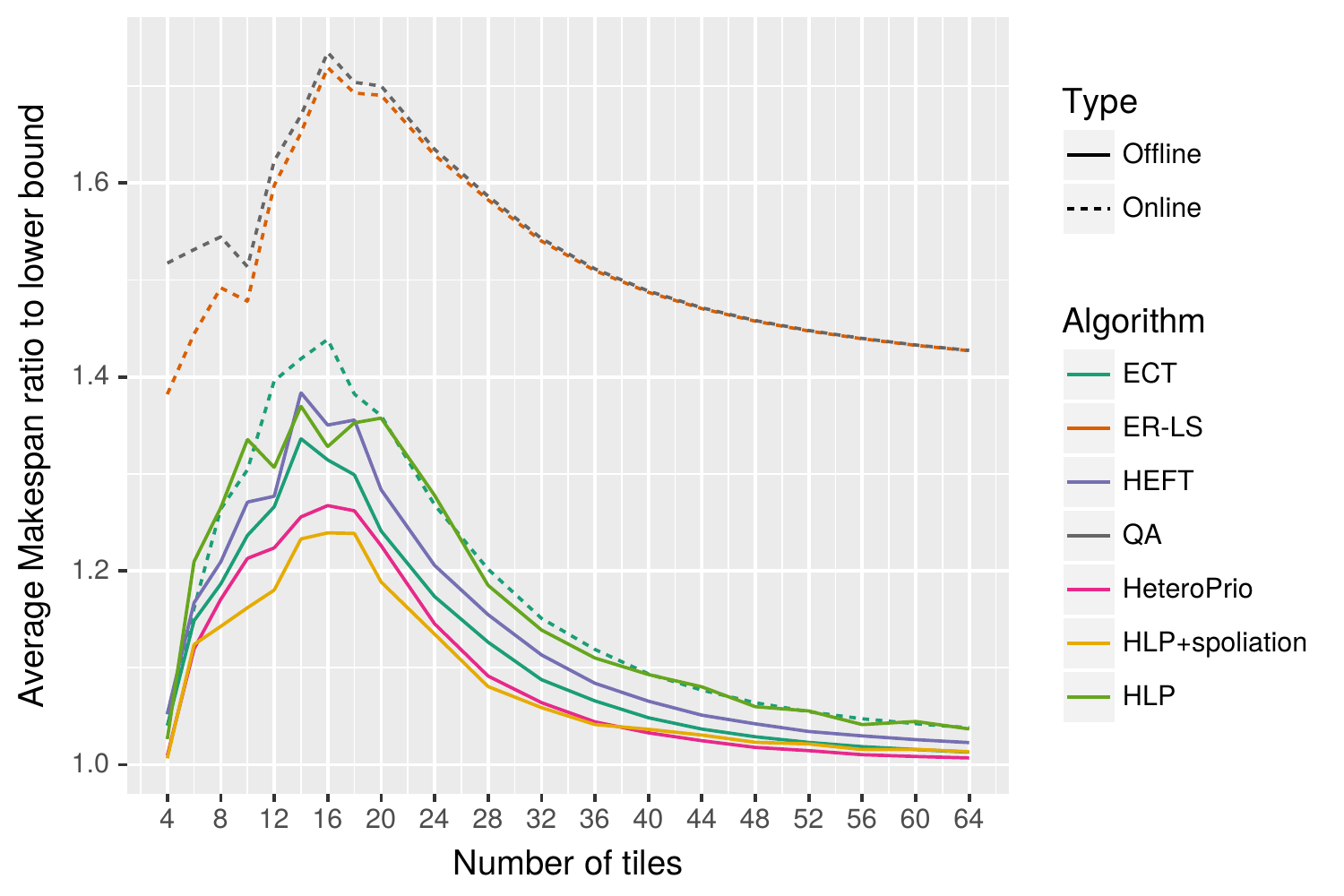}
    
    \includegraphics[width=\textwidth]{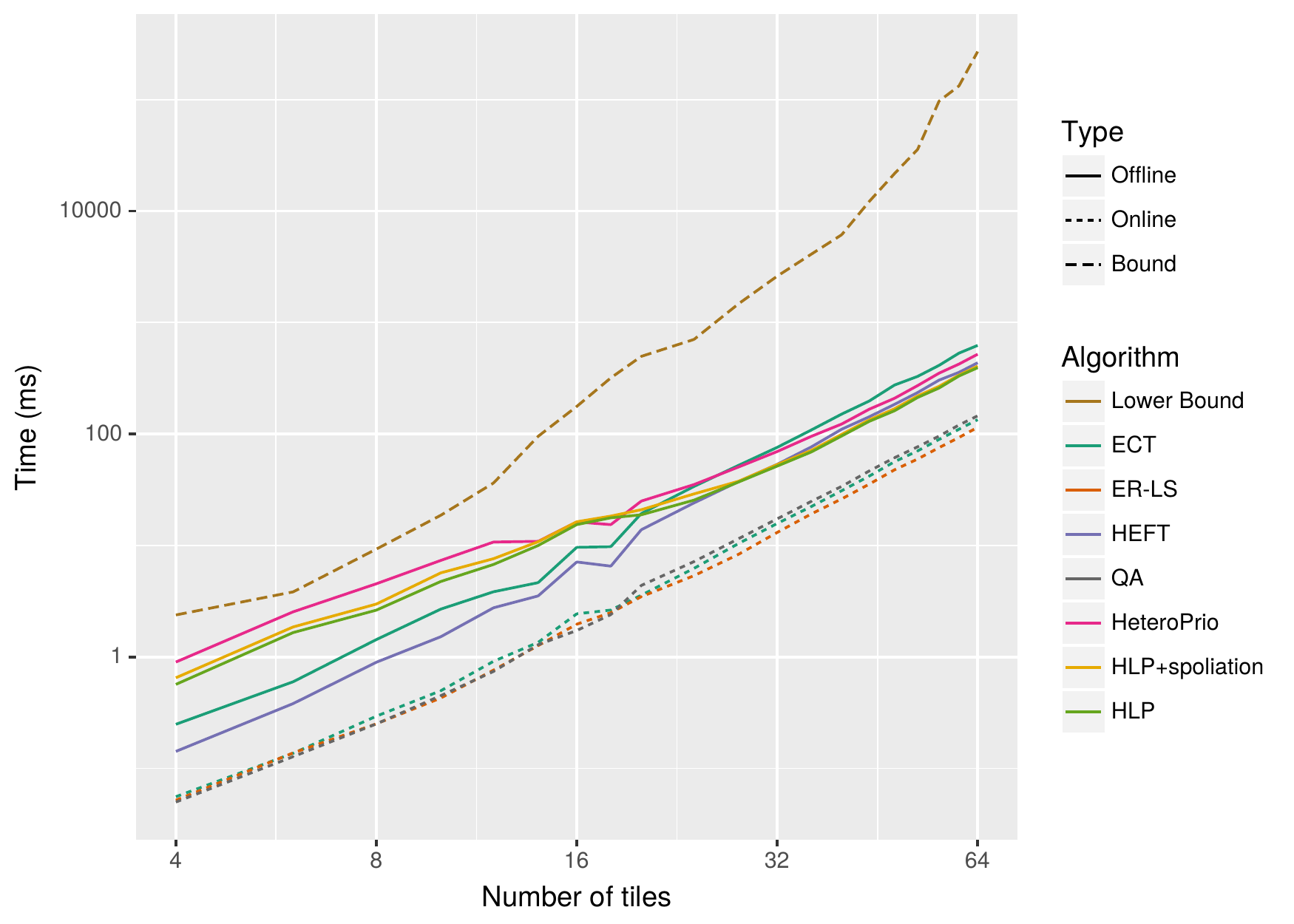}
  \end{center}
  \caption{Normalized performance and computation times for Linear
    Algebra with dependencies, averaged over different numbers of
    processors and different applications.}
  \label{fig:expe:bordeaux:time}
\end{figure}

\Cref{fig:expe:bordeaux} depicts the corresponding results.
These plots show the results for each individual instance, where each column corresponds to a different application, and each row corresponds to a different amount of resources.
In all plots, the $x$ axis shows the size of the matrix (expressed as number of tiles).
These graphs are hard to interpret, so we also provide an average view in the top plot of Figure~\ref{fig:expe:bordeaux:time}, where all results for each matrix size are averaged over the different applications and platform sizes.
This allows us to make the following observations:

\begin{itemize}
\item With this type of instances, scheduling becomes easier when the number of tiles is very small or very large.
Indeed, when the number of tiles is small, the graph is very small and simply scheduling (almost) all tasks close to the critical path on the GPU is enough to achieve low makespan.
On another hand, when the number of tiles is large, the overall work is dominated by a large number of a particular type of tasks (matrix products in many cases), so that the area bound dominates the schedule length.
Of course, the middle ground depends on the number of resources: from 8 to 20 tiles on a small platform, and from 16 to 32 on a large platform.
\item Concerning on-line algorithms, only the ECT algorithm has a ``reasonable'' behavior for large size instances: the two other algorithms behave very similarly and their performance does not converge to an almost optimal one when the problem size becomes large.
The reason is that these algorithms target a given approximation ratio, without trying to obtain a better solution if available.
The comparatively better performance of ECT also explains why we do not include Mixed-ECT-QA in the results: it actually would achieve the same results as ECT, since there is no reason to switch to another algorithm.
\item Spoliation does improve the performance of the HLP algorithm on these instances, and makes it the best performing algorithm.
However, as we will see later, this comes at the price of a very high computational cost.
\item Among the low-cost heuristics, HeteroPrio achieves the best results, followed by Off-line ECT and HEFT.
The difficult instances for HeteroPrio are when the numbers of CPUs and GPUs are close, because there are fewer opportunities for spoliation.
\end{itemize}

\Cref{fig:expe:bordeaux:time} presents the running times of all algorithms, gathered and averaged over all the instances with the same number of tiles, and with a logarithmic scale on both axes.
We see that all on-line algorithms have the same computational cost, and the same observation holds for off-line algorithms.
On all instances, the number of tasks is of the order of the cube of the number of tiles, and we indeed see a polynomial dependency on the graph.
The plot also includes the time to compute the solution of the linear program used to obtain the lower bound, and whose solution is also used for the assignment of HLP.
This part of the computation however has a much higher computational cost, which needs to be taken into account when analyzing the results of HLP.

\section{Conclusion}\label{sec:conclusion}

\subsection{Synthesis}

We presented a comprehensive survey with the objective of questioning how to efficiently schedule the tasks of a parallel application on hybrid computing platforms composed of multi-cores and accelerators.
A first and obvious output of this study is to provide a clear synthesis of all the existing algorithms addressing this question.
We have revisited all published algorithms in the area within the same unified framework.
They have been presented with both an emphasis on their key underlying ideas and with a systematic analysis of the theoretical worst-case performance.
Another output is the design of a common testbed for comparing the various algorithms, both in terms of the quality of resulting schedule and in terms of the actual running time to compute it.
The benchmark used to perform the comparison is composed of a variety of random instances and several realistic data extracted from actual applications.
We encourage the community to use this benchmark for further investigations.

\subsection{Lessons learned}

There are several lessons that can be learned from this study.
First, and not surprisingly, there is no straightforward conclusion in term of determining what is the best scheduling policy whatever the instances.
The choice of an algorithm is always a matter of trade-offs.
Second, we showed that the problem of designing generic scheduling on hybrid parallel platforms is tractable, assuming a reasonably simple computational model.
This survey should be considered as a useful study that provides solid arguments to the users of such platforms.
On the practical side, the old, cheap and robust HEFT algorithm still behaves well.
It is a good competitor but it does not have theoretical guaranties that would prevent too bad executions on some instances.

\subsection{Extensions}\label{sec:extensions}

This study provides a full picture of the existing scheduling algorithms for hybrid platforms under the restricted assumptions that correspond to today's platforms. There are several research directions for extending the algorithms.

The first generalization is to consider the case with $K > 2$ types of computing components, for which several PTAS have been designed for the case of independent tasks~\cite{bonifaci2012a, gehrke2016a}.
We believe that many algorithms presented in this paper can be adapted for these new problems and most of them will keep constant approximation guaranties (depending linearly on the number of processor types, as it is the case for HLP~\cite{concurrency}).
Even though they are still rare, we can envision the development of many devices dedicated to specific use, such as TPUs or FPGAs.
The setting with more than one type of accelerator is therefore expected to become of practical interest soon.

A second direction is to extend the model of sequential tasks to parallel tasks. A first attempt has been proposed by considering moldable executions on the CPU part~\cite{BleuseIEEE}, which does not change significantly the approximation results.
Another challenging extension is to take into account the communication cost and the congestion on the network between CPUs and GPUs that has been neglected in most existing algorithms, except for an extension of HLP proposed by Aba et al.~\cite{LPcommunications}.
The problem for obtaining useful results under any communication model is that the analysis are closely related to the underlying architecture, and are therefore hard to generalize.
Finally, a $(2+\alpha)$-dual approximation has been
proposed~\cite{bleuse2014a} to take into account affinity scores
between tasks and processors that may represent data locality. Other
approaches to tackle such locality issues may lead to lower
approximation ratios.

\bibliographystyle{plain}

\end{document}